\documentclass[fleqn,10pt]{wlscirep}
\usepackage[utf8]{inputenc}
\usepackage{amsfonts}
\usepackage{amsmath}
\DeclareMathOperator*{\argmax}{arg\,max}

\usepackage[T1]{fontenc}
\usepackage{amssymb}
\usepackage{eufrak}
\usepackage{stackrel}
\usepackage{tikz-cd}
\makeatother
\title{Benchmarking Inverse Optimization Algorithms for Materials Design}
\author[1]{Hanfeng Zhai\footnote{Email: \tt hz253@cornell.edu}\footnote{Present Address: Department of Mechanical Engineering, Stanford University}}
\author[2]{Hongxia Hao\footnote{Email: \tt hongxiahao@microsoft.com}}
\author[1]{Jingjie Yeo\footnote{Email: \tt jingjieyeo@cornell.edu}}
\affil[1]{Sibley School of Mechanical and Aerospace Engineering, Cornell University, Ithaca, New York 14850, USA}
\affil[2]{Microsoft Research AI4Science, Beijing, China}

\begin{abstract}
Machine learning-based inverse materials discovery has attracted enormous attention recently due to its flexibility in dealing with black box models. Yet, many metaheuristic algorithms are not as widely applied to materials discovery applications as machine learning methods. There are ongoing challenges in applying different optimization algorithms to discover crystals with single- or multi-elemental compositions and how these algorithms differ in mining the ideal materials. We comprehensively compare 11 different optimization algorithms for the design of single- and multi-elemental crystals with targeted properties. By maximizing the bulk modulus and minimizing the Fermi energy through perturbing the parameterized elemental composition representations, we estimated the unique counts of elemental compositions, mean density scan of the objectives space, mean objectives, and frequency distributed over the materials' representations and objectives. We found that nature-inspired algorithms contain more uncertainties in the defined elemental composition design tasks, which correspond to their dependency on multiple hyperparameters. Runge Kutta optimization (RUN) exhibits higher mean objectives whereas Bayesian optimization (BO) displayed low mean objectives compared with other methods. Combined with materials count and density scan, we propose that BO strives to approximate a more accurate surrogate of the design space by sampling more elemental compositions and hence have lower mean objectives, yet RUN will repeatedly sample the targeted elemental compositions with higher objective values. Our work shed light on the automated digital design of crystals with single- and multi-elemental compositions and is expected to elicit future studies on materials optimization such as composite and alloy design based on specific desired properties.

\end{abstract}
\begin{document}

\flushbottom
\maketitle
%
%
\thispagestyle{empty}


\section*{Highlights}

\begin{itemize}
    \item Present a testbed that can implement many state-of-the-art optimization algorithms, and benchmarked 11 different optimizations spanning from Bayesian optimization to genetic algorithms, particle swarm optimization, etc. by evaluating multiple proposed metrics.
    \item Hybrid grey wolf optimization, hybrid improved blue whale optimization, and genetic algorithms generally display higher variances from repeated experiments for the unique elemental count, objective space density scan, and mean objective values. Bayesian optimization and Runge Kutta optimization generally show lower variances. 
    \item Runge Kutta optimization exhibits higher mean objective values and Bayesian optimization shows lower mean objectives, compared with other optimizations benchmarked.
    \item By combining estimations from materials count and normalized frequency, one deduces Bayesian optimization aims to approximate a more accurate surrogate of the design space and hence sample more elemental compositions, whereas Runge Kutta optimization tends to repeatedly sample the discovered optimal materials multiple times.
\end{itemize}

\section{Introduction}

Designing materials with desired tailored properties has always been a key goal in human society, from ancient metallurgy to modern nanotechnology. Traditionally, most research conducted in the field of designing crystals with multi-elemental compositions relies on costly experimental synthesis and time-consuming trial-and-error processes. Advances in computational engineering can greatly accelerate materials design efficiency. To achieve the design goal, many optimization methods were developed for the design process. These methods can be categorized as first and zeroth orders, where first-order optimization methods are gradient-based methods, such as gradient descent and Newton-Raphson, that rely on the problem being first-order differentiable. First-order methods are widely applied to topology optimization coupled with finite element methods for mechanical design, including classical problems like the design of beams\cite{beam_matlab_88lines} and trusses\cite{truss_design_cmame}. 
However, the primary drawback of first-order optimization is that the differential form of the problems may not exist, or be easily obtained, in many (or most) real-world application scenarios. In many cases, the governing equations for the problem do not exist as differentiable forms. Moreover, first-order optimization tends to get ``trapped'' in local optima close to the initial design which heavily relies on prior physical understanding of the problem to ameliorate. Hence, metaheuristic optimization methods, such as genetic algorithms and particle swarm optimization, are more flexible and adaptive to complex problem domains. These characteristics enable metaheuristics to be applicable to multiscale materials design problems, such as molecular simulations, machine learning surrogates, and density functional theory calculations. For example, molecular dynamics (MD) simulations model the interactions of particles based on interatomic potentials\cite{interatomic_graphene_zhai, JJ_Carbon_graphene_defect_thermal}, DFT simulations calculate the energy based on pseudo-potentials for the Hamiltonian\cite{dft_pbtio3_benedek_prl}, and ML surrogates represent the material's structure-property map with a neural network\cite{Buehler_ml_surrogate}. In these situations, obtaining first-order derivatives is either an extremely taxing task or may even be intractable. To overcome these challenges, recent literature highlights the utility of metaheuristics for elemental composition design, in contrast to gradient-based optimizations. For example, Ramprasad and coworkers developed genetic algorithm methods coupled with deep learning surrogate models for designing polymer dielectrics\cite{rampi_polymer_dielectric_ga} and used similar strategies to design polymers with high glass transition temperatures and bandgaps\cite{rampi_ga_polymer_design}. Winter et al. combined machine learning surrogate with particle swarm optimization (PSO) to design drugs\cite{pso_molecule_1}. Weiel et al. used PSO to select parameters in molecular dynamics simulations\cite{pso_md_parameter}.

Many numerical methods exist for design optimization and two of the most widely applied methods are Bayesian optimization (BO) and deep reinforcement learning (DRL). BO leverages Gaussian process regression to fit a surrogate model of the design space and leverages Bayesian statistics to iteratively improve the accuracy of such a surrogate by sampling more points via an acquisition function. BO is frequently applied in materials design. For example, Shin et al. combined micromechanical experiments and BO to design spiderweb nanomechanical resonators\cite{miguel_bo}. Chen and coworkers developed a series of BO frameworks for mechanical design and elemental composition design using finite element simulations\cite{chen_fem_bo}, molecular simulations\cite{bo_md_design_solarcells, weichen_des_mo_bo}, and machine learning surrogate models\cite{chen_bo_design_ml_model, chen_bo_ml_2}. Zhai and Yeo combined BO and individual-based modeling of bacteria to design antimicrobial nanosurfaces\cite{bo_biofilm}. In another popular method, DRL utilizes a deep neural network (DNN) as an agent to interact with the environment to receive rewards from different output properties and learn the best strategy to achieve the design goal. When applying DRL for materials design, the evaluations of the material properties are treated as the environment for the agent to learn the correct policy to achieve the design goal. For instance, Gu and coworkers developed a DRL framework for designing tough graphene oxide\cite{gu_drl_graphene} and continuum composite materials\cite{gu_drl_comp_mat}. Farimani and coworkers developed DRL frameworks for porous graphene for water desalination\cite{amir_drl_graphene_water}. 

Can these optimization methods be uniformly applied to the same problem and how do they differ? To answer these questions, benchmark studies are urgently needed. One particular reason for the urgency is efficiency. Optimization methods are intended to improve the efficiency of the design problems. However, when companies or research groups want to employ design optimization for their problems of interest, it is of utmost importance to select optimization methods that are best suited to their specific problems. Selecting ideal optimization methods from the massive amounts of existing literature is non-trivial: one may fall into the ``trial-and-error'' trap and the efficiency of solving the problem is significantly reduced. Moreover, ML-based studies for materials design are often focused on either large-scale molecular structures (i.e., polymers) or continuum structures. This is in part because it is easier to extract structure-property relationships for larger structural bodies in the context of continuum modeling, where first-order derivatives are easy to obtain. Specifically for materials composed of multiple elements, many studies leverage molecular graphs or use natural language processing to find this structure-property map. In contrast, while there are some studies on the design of small molecules using language representation with generative models\cite{nlp_small_molecule_design}, a huge gap exists in applying general optimization methods for designing such multi-elemental compositions. Therefore, following our previous discussion, the specific problem to be posed herein is how different optimization methods differ in the context of design optimization.



\begin{figure}[htbp]
    \centering
    \includegraphics[scale=0.33]{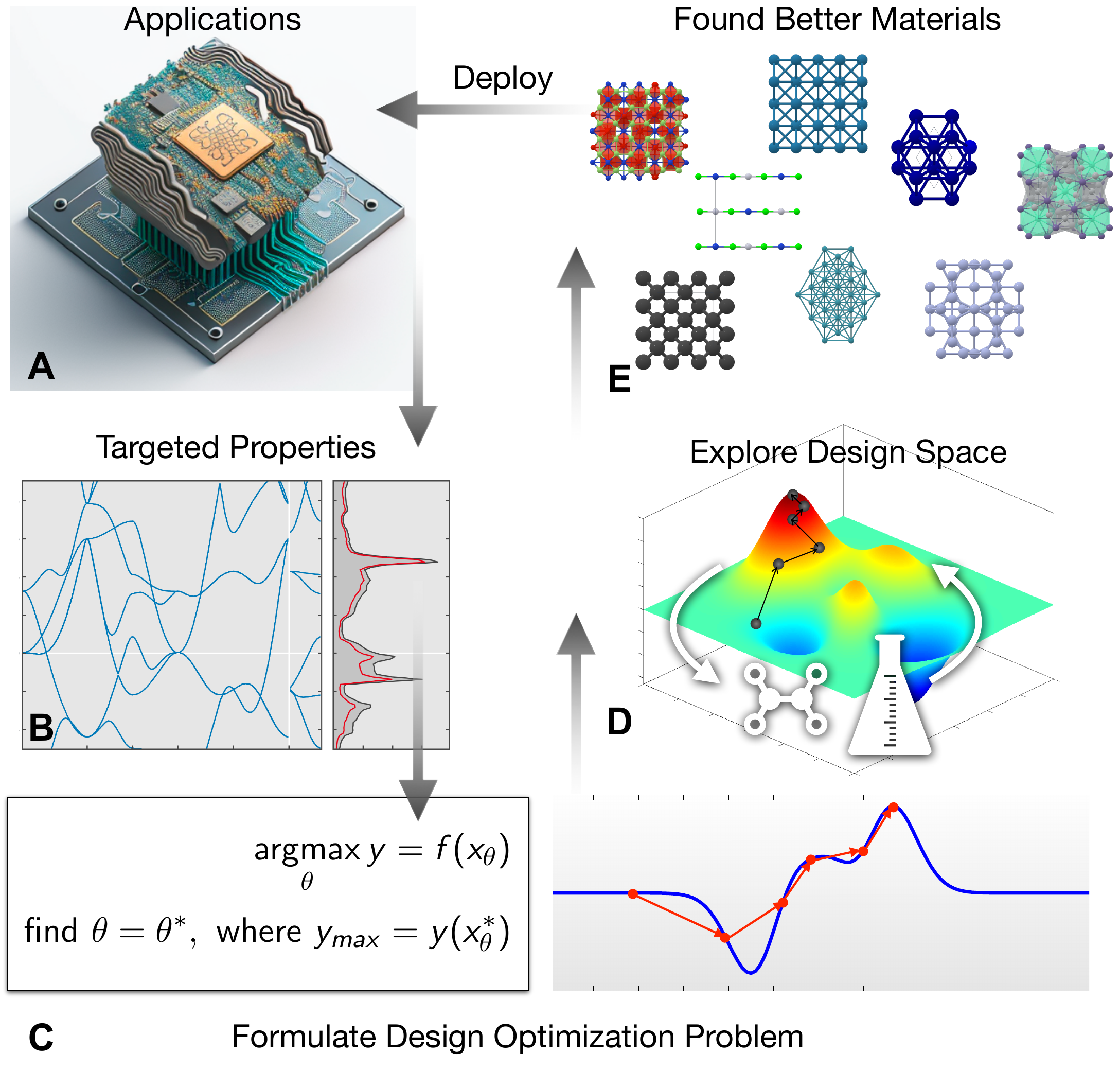}
    \caption{The general overview of the materials design process. {\bf(A)} Schematic for defining the intended industrial application. {\bf(B)} The process of identifying the desired material properties such as high bulk modulus and low Fermi energy. {\bf(C)} The schematic for formulating a digital design optimization problem using numerical models to discover or design elemental composition with the desired properties. {\bf(D)} Different optimization methods use exploration/exploitation strategies to generate a design space for exploring the best materials. {\bf(E)} The optimal elemental compositions are extracted from the design space exploration and are ready for real-world applications. In this study, we focus on formulating an optimization framework that enables the application of the most current popular optimization methods to search for desired elemental compositions and compare different optimizations based on specific metrics. Finally, the ``mined'' elemental compositions are analyzed.}
    \label{fig1}
\end{figure}

To address these challenges, we propose and conduct a benchmark study by designing a framework to implement most state-of-the-art optimization methods. We benchmark many widely employed metaheuristic optimizations, including Bayesian optimization, genetic algorithms, particle swarm optimization, and simulated annealing. We further compared novel optimization methods, such as the arithmetic optimization algorithm and Runge Kutta optimization, for a total of 11 different optimization algorithms. We took advantage of the large numbers of open-source software for computational elemental composition design, including pymatgen\cite{pymatgen}, Materials Project\cite{mat_proj}, matminer\cite{matminer}, and mealpy\cite{mealpy}. The material evaluations are enabled by surrogate models: pretrained graph neural network models to predict the elemental compositions' properties based on input graph structure. The general schematic of our study is shown in Figure \ref{fig1}. We begin with inspiration from the general premise of industrial or fundamental research problems: one needs to design elemental compositions based on specific applications. Given this application, the targeted physical properties must be identified based on the design requirements, such as high toughness or low thermal conductivity. Based on these targeted applications and properties, the design problem can then be formulated mathematically together with the corresponding optimization algorithms. The design space can be explored to extract the optimal elemental compositions from the materials database(s). Here, we will mainly focus on the optimization process (Figure \ref{fig1} \textbf{C}, \textbf{D}, \& \textbf{E}) to propose simplified target material properties and focus on the algorithmic implementations and benchmarking. {The significance of this work lies in that there is a lack of studies that evaluate the performance of different optimization methods, in contrast with the abundance of published literature that applies specific optimization methods for tailored materials design tasks. Our study may benefit future research in ways such as: (1) Researchers can select different algorithms that correspond well with their potential applications based on our evaluation results; (2) Researchers can use or refer to our framework and benchmark other optimization algorithms of their interests. Our major contributions are to (1) propose a framework that can implement different algorithms, which generalizes the materials' inverse design tasks, (2) evaluate different optimization methods based on defined metrics, and (3) highlight ideal algorithms for different potential applications based on our evaluations. Some results were unexpected: the commonly employed BO did not stand out in terms of the mean objective evaluation, while surprisingly RUN outperformed on both the stability and objective evaluations.}

This paper is articulated as follows. We present an overview of the underlying methodology, describe the workflow of connecting machine-learned models of materials properties with different optimization methods, and benchmark the results obtained from 11 different optimization algorithms in Section \ref{section_results}. We then discuss the algorithms that are developed to perform the automated optimizations in Section \ref{section_discussion}. A detailed explanation of the methodologies employed is presented in Section \ref{sec_method} and further information is contained in the ESI.  

\begin{figure}[htbp]
    \centering
    \includegraphics[width=0.99\textwidth]{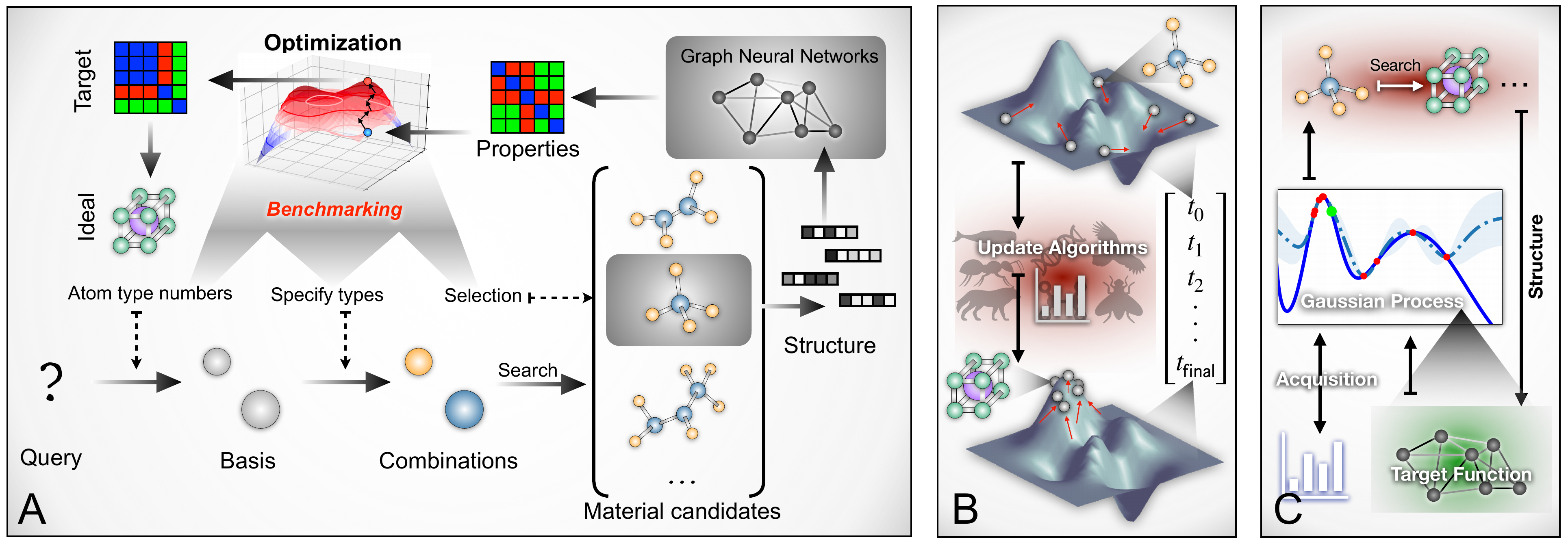}
    \caption{Schematic for optimization benchmarking and different optimization algorithms employed in the study. {\bf(A)} Workflow for discovering elemental compositions by combining graph neural network (GNN) surrogate models with defined design variable randomization to select optimal elemental compositions with targeted material properties. See Section \ref{sec_method} \& ESI for details. {\bf(B)} Schematic for optimization through metaheuristic algorithms, starting with a set of random particles (or population) sampled in the design space, to iteratively update and eventually cluster around globally optimal values, indicating the search for elemental compositions with desired properties. {\bf(C)} Bayesian optimization methods for elemental composition discovery, seeking to build a surrogate model of the design space by employing Gaussian process regression and searching for the next material evaluations based on an acquisition function. Optimal elemental compositions can be extracted from the design space surrogate.}
    \label{fig2}
\end{figure}

\section{Methods\label{sec_method}}

\subsection{Inverse Optimization Framework}


\subsubsection{Problem Formulation}

Our goal is to benchmark design optimization algorithms for inverse elemental composition design, in which we formulate the problem as maximizing an objective function corresponding to materials properties. We benchmark using a model problem of maximizing the bulk modulus while simultaneously minimizing the Fermi energy. While not specific to any applications, our prototype model is useful for benchmarking and unveiling characteristics of different optimization algorithms for elemental composition design and discovery. To maximize an objective function, $\mathcal{J}$, as a single-objective optimization problem which is defined as the difference between the bulk modulus and the Fermi energy, $\mathcal{J} = K-E_{Fermi}$:
\begin{equation}
        \begin{aligned}
            \argmax_{n_{atom},\ \xi_n,\ \eta} \mathcal{J}=K-E_{Fermi},\\
             {\rm where}\quad K, E_{Fermi} = MEGNet(\mathcal{G}_\Theta),\\\rightarrow \mathcal{G}= \Omega(n_{atom},\ \xi_n,\ \eta);\  \Theta = [n_{atom}, \xi_n, \eta]\\{\rm subject\ to}\quad n_{atom} \in [1,4] \ {\rm or} \ \equiv 1,\ \xi_n \in [0,100],\ \eta\in[0,100]
            \label{eqn_opt_problem}
        \end{aligned}
    \end{equation}Here, the problem is defined as a discrete constrained optimization problem. The input graph structure $\mathcal{G}$ is related to the properties $K$ and $E_{\rm Fermi}$, which can be considered as a forward mapping. The forward mapping of fast properties predictions is enabled based on the graph neural network structure MatErials Graph Network (MEGNet) \cite{megnet}. Further details are provided in Section \ref{sec_method} \& ESI.

The optimization problem can be solved using various optimization algorithms. Here, we explored 11 different state-of-the-art optimizations to conduct benchmarking for materials discovery. The methods include Bayesian Optimization\cite{method_bo}, Deep Reinforcement Learning\cite{sutton2018reinforcement}, Genetic Algorithm\cite{GA_optimization}, Particle Swarm Optimization\cite{PSO_optimization}, Hybrid Grey Wolf Optimization\cite{GWO_optimization}, (Hybrid-Improved) Whale Optimization Algorithm\cite{hiwoa_optimization}, Ant Colony Optimization\cite{ACO_optimization}, Differential Evolution\cite{DE_optimization}, Simulated Annealing\cite{sa_optimization}, Bees Optimization\cite{Bees2_optimization}, Arithmetic Optimization\cite{AOA_optimization}, and Runge Kutta (RUN) Optimization\cite{RUN_optimization}. These algorithms fall into two main categories: machine learning-based optimizations and metaheuristic optimizations (Figure \ref{fig2} {\bf B} \& {\bf C}). We use different strategies to perturb $\Theta$ to maximize $\mathcal{J}$, which will be elaborated in Section \ref{sec_method} \& ESI. Here, we consider single and multi-element composition optimization scenarios. For single-element composition optimization, the design variable $\Theta$ is restricted to 1, and there are only two design variables $\xi_n$ and $\eta$. For multi-element composition optimization, the screened elemental compositions contain $n_{atom} \in [1,4]$. The materials libraries as indicated in Figure \ref{fig2} consist of a series of MP IDs and the randomly generated $\eta$ is used to select the corresponding MP ID.


Following the discussion from Equation (\ref{eqn_opt_problem}), the optimization is benchmarked and connected through the material evaluations. We utilize MEGNet for materials properties evaluation mainly for two considerations: (1) Fast inference. By using the pretrained MEGNet model ({\sf mp-2019.4.1}), the properties can be obtained ``on the fly'' given the input graph structure, getting rid of the lengthy computational process via traditional numerical methods like molecular dynamics (MD) or density functional theory (DFT). (2) Better screen-ability of the design space. Given the pretrained MEGNet model, one can infer the corresponding properties given generated input graph $\mathcal{G}$ and hence allowing the evaluation of a large amount of ``untouched'' materials. On the other hand, due to the lack of existing data for the properties of many potential materials, one may not directly obtain the properties given the input elements just from a query. Also, due to the lack of pseudo- and/or interatomic potentials, MD or DFT calculations may not be conducted. {Here, we employ MEGNet due to its reported accuracy and stability. As reported by Chen et al.\cite{megnet}, the MAE for the prediction for the Fermi energy and bulk modulus are around 0.028 and 0.05, respectively. MEGNet is not the only option for materials properties inference, other existing frameworks with competitive performance include CGCNN\cite{MODNet_paper}, MODNet\cite{MODNet_paper}, H-CLMP(T)\cite{H-CLMP(T)}, and Crystal graph attention networks\cite{Crystal_graph_attention_networks}. Since our focus is on the optimization methods, we do not emphasize the surrogate model evaluations. Notably, many recent works focus on materials property prediction benchmarks\cite{Fung_GNN_benchmark, material_prediction_benchmark}.}

For this problem, the inverse optimization problem is to directly perturb the graph structure such that the materials' properties can be tailored. However, it is difficult to represent different materials' graph structures containing different elemental numbers in the same dimensions for optimization. Hence, we parameterize the graph structure $\mathcal{G} = \mathcal{G}_\Theta$ so that the algorithms tune the design variables $\Theta = [n_{atom}, \xi_n, \eta]$ to change the elemental composition $\mathcal{G}_\Theta$. Here, as visualized in Figure \ref{fig2} \textbf{A}, the optimization begins with variable $n_{atom}$, as the number representing different types of elements in the elemental compositions. We may denote the different atoms in the elemental compositions as the {\em atomic dimensions}. Follows is a random number distributed in $[0,100]$ as $\xi_n$ to assign atomic types to the atomic dimensions, which we may define as the {\em material basis}. For example, if the atomic dimension $n_{atom}=3$ we may obtain an element basis from $\xi_n$ like \{C-H-O\}. Given this material basis, one can generate a list of potential elemental compositions from the {\em Materials Project}, for which we denote as the {\em materials library}. One can then use $\eta$ to select the candidate elemental compositions from the materials library. If following our previous example, the materials library may look like \{$\rm CH_3O$, $\rm C_2H_6O$, $\rm C_3H_6O$, ...\} and our selected material may be $\rm C_2H_6O$. For how $\xi_n$ and $\eta$ determine the atomic graph structure $\mathcal{G}$ in the algorithm, please refer to {\em Extension on Problem Formulation} in ESI.

\subsubsection{Workflow Automation}

One of the main contributions of our work is the development of an optimization framework that is flexible for many implementations of state-of-the-art optimization methods. The key point in such a framework is the automation of the optimization, where the update of the next set and the current set of the materials evaluation are connected. The automation is enabled through the graph neural network surrogate model as formulated in Equation (\ref{eqn_opt_problem}), the MEGNet. The whole workflow can be simplified to the expression:\begin{equation}
\underbrace{(\overbrace{MEGNet(\Theta_{n+1})}^{\text{evaluation}}\leftarrow\Theta_{n+1})}_{\text{exploration}} \overset{\text{search}}{\underset{\text{feedback}}{\leftrightharpoons}}\underbrace{IOA({\Theta}_n,\overbrace{MEGNet({\Theta}_n)}^{\text{evaluation}};\mathbf{P}_{IOA})}_{\text{exploitation} }\label{eqn_opt_automation}
\end{equation}where $IOA$ stands for ``inverse optimization algorithms''. Assumes the algorithm evaluates the $n^{th}$ set of elemental compositions with the design variable $\Theta_n$, obtaining the corresponding properties $MEGNet(\Theta_n)$, by employing the hyperparameters $\mathbf{P}_{IOA}$ specified by from different optimization methods, one then obtain the design variables for the next set of evaluation $\Theta_{n+1}$, for which the evaluated properties $MEGNet(\Theta_{n+1})$ then feeds back to the IOA and iteratively search for the next set. This loop allows one to get rid of the traditional ``trial-and-error'' approach and only need to set the hyperparameters to obtain the optimal elemental compositions given the targeted properties.

\begin{figure}
    \centering
    \includegraphics[width=.9\linewidth]{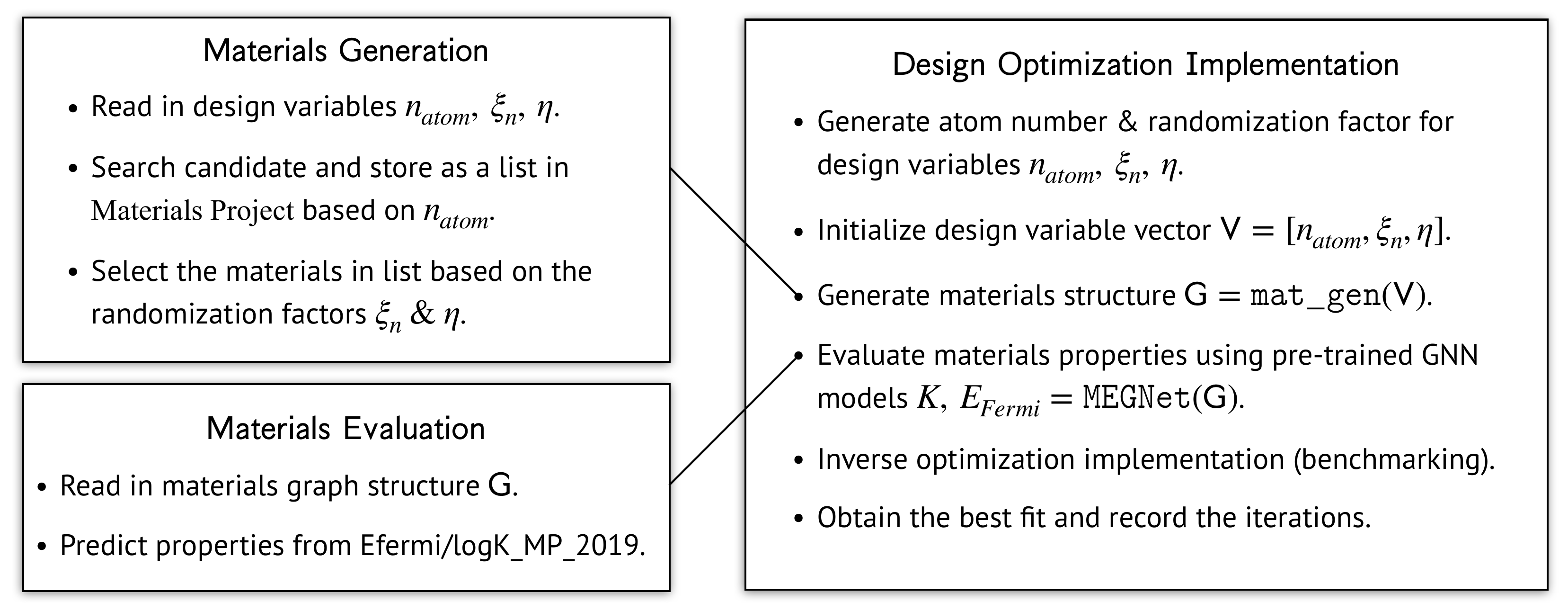}
    \caption{Algorithmic flow chart for the optimization framework, corresponding to Equations (\ref{eqn_opt_problem}) \& (\ref{eqn_opt_automation}) and Figure \ref{fig2}. }
    \label{fig_meth_flow-chart_matopt}
\end{figure}

{ The algorithmic flowchart is visualized in Figure \ref{fig_meth_flow-chart_matopt} (Equations (\ref{eqn_opt_problem}) \& (\ref{eqn_opt_automation})). The formulated basic optimization method is shown in the right block, while the materials generation and evaluation processes are laid out in the left blocks. In our approach, the algorithm is initialized and repeated based on different random seeds for the five attempts in Figure \ref{fig4}. The hyperparameters for each algorithm are provided in the ESI and were selected from commonly used default values for each algorithm. }


\subsection{Optimization Algorithms}

\subsubsection{Bayesian Optimization}
The core of machine learning (ML)--based optimization algorithms is to employ the explored materials data to fit a surrogate assisting the optimization process. Here, there are two main ways of adopting ML methods for design optimization: (1) building up a surrogate model of the design space for final exploration, namely the Bayesian optimization method; (2) utilizing a deep neural network (DNN) to learn the interactions of input elemental compositions and their corresponding material properties, which can be considered as the interaction of actions and their environment in which this DNN is considered as an ``agent''. This method is known as deep reinforcement learning. The detailed mathematical formulation and derivations are shown in the {\em Machine Learning Based Algorithms} \& {\em Metaheuristic Optimization} in ESI.

For approach (1), following the formulated problem in Equation (\ref{eqn_opt_problem}), we now assume the optimization algorithm is initiated by a set of $m$ materials evaluations from MEGNet: $K_1$, $K_2$, $K_3$, $..., K_m$, $(E_{\rm Fermi})_1$, $(E_{\rm Fermi})_2$, $(E_{\rm Fermi})_3$, $..., (E_{\rm Fermi})_m$ = $MEGNet(\Omega(\Theta_1), \Omega(\Theta_2), ..., \Omega(\Theta_m))$; one then obtains data lies in the mapping of $M: \Theta\rightarrow\mathcal{J}$, in which we assume $M$ exist and continuous. Here, this $M$ represents the design space, i.e., the projection from the input design variables to the objectives. $M$ can be approximated using Gaussian process regression (GPR) from the generated data in the materials evaluation, which can be simplified to the form:\begin{equation}
    \begin{aligned}
        \left(\mathcal{J}_1, \mathcal{J}_2, \mathcal{J}_3,...,\mathcal{J}_m\right) = GP(\Theta_1, \Theta_2, ...,\Theta_m; \mathbf{p}_{GP})
    \end{aligned}
\end{equation}where $\mathbf{p}_{GP}$ are all the hyperparameters involved in setting up the regression model, elaborated in detail in the ESI. Given a surrogate model of the design space built from GPR, one can improve the accuracy of such a surrogate by iterative sampling more points. The core of BO here would be the strategies to sample the next set of material evaluations from $\Theta$.

The next set of design variables $\Theta_{next}$ is selected via maximizing the acquisition function $\mathcal{A}$, writes:\begin{equation}
    \Theta_{next} = \argmax\mathcal{A} (\Theta_1, ..., \Theta_m, \mathcal{J}_1,..., \mathcal{J}_m; \mathbf{p}_{Aq})
\end{equation}where we use the upper confidence bound as the acquisition function (see {\em Bayesian Optimization} in ESI). BO utilizes GPR and acquisition function to efficiently sample the design space for a more accurate approximation towards the formulated optimization goal ($\max\mathcal{J}$).

\subsubsection{Metaheuristic Optimizations}

Metaheuristic optimization is a family of optimization techniques that are used to find the optimal solution by exploring a large search space. They are often inspired by natural processes or social behavior, such as animal behavior (e.g., ant, whale, wolf), physical process (i.e., annealing of metals), biological processes (i.e., gene mutation), or fundamental mathematics.

Metaheuristic algorithms are typically applied to complex optimization problems where traditional methods are either not feasible or inefficient. They are highly flexible and can be adapted to a wide range of problem domains, in which elemental composition design is our target scenario. Metaheuristics are often considered "black box" optimization methods, as they do not require knowledge of the problem structure. More specifically, they do not require the existence of the first-order differentiation form of the governing equations of the problem.


If we assume the simplest form, one can simplify and interpret the metaheuristic optimization process by sampling a set of random particles in the design space and the optimization process refers to the update of these sampled particles (For the generalization of this form see {\em Methodology Generalization} in ESI). The update rules may be inspired by different natural processes, corresponding to the optimization methods employed. If one assumes there are a total of $d$ dimension of the problem domain, and denotes the rule as $\mathcal{R}$, the update rules for the $i^{th}$ particle write:\begin{equation}
    X_{i}^d \leftarrow X_{i}^d + \mathcal{R} ( X_{i}^d, X_{j}^d,\mathbf{p}_{MH})\label{eqn_meta_update}
\end{equation}where $X$ is the location of the particle, and the $i^{th}$ particle's location may be related to it adjacent particle(s) $j$. $\mathbf{p}_{MH}$ denotes the hyperparameters employed in the metaheuristic optimizations. We here only present the generalized form to identify the key essence of metaheuristics (details of each algorithm see {\em Metaheuristic Optimization} in ESI).

\section{Results\label{section_results}}



\subsection{Elemental Composition Design Benchmarks}

\begin{figure}[htbp]
    \centering
    \includegraphics[scale=0.32]{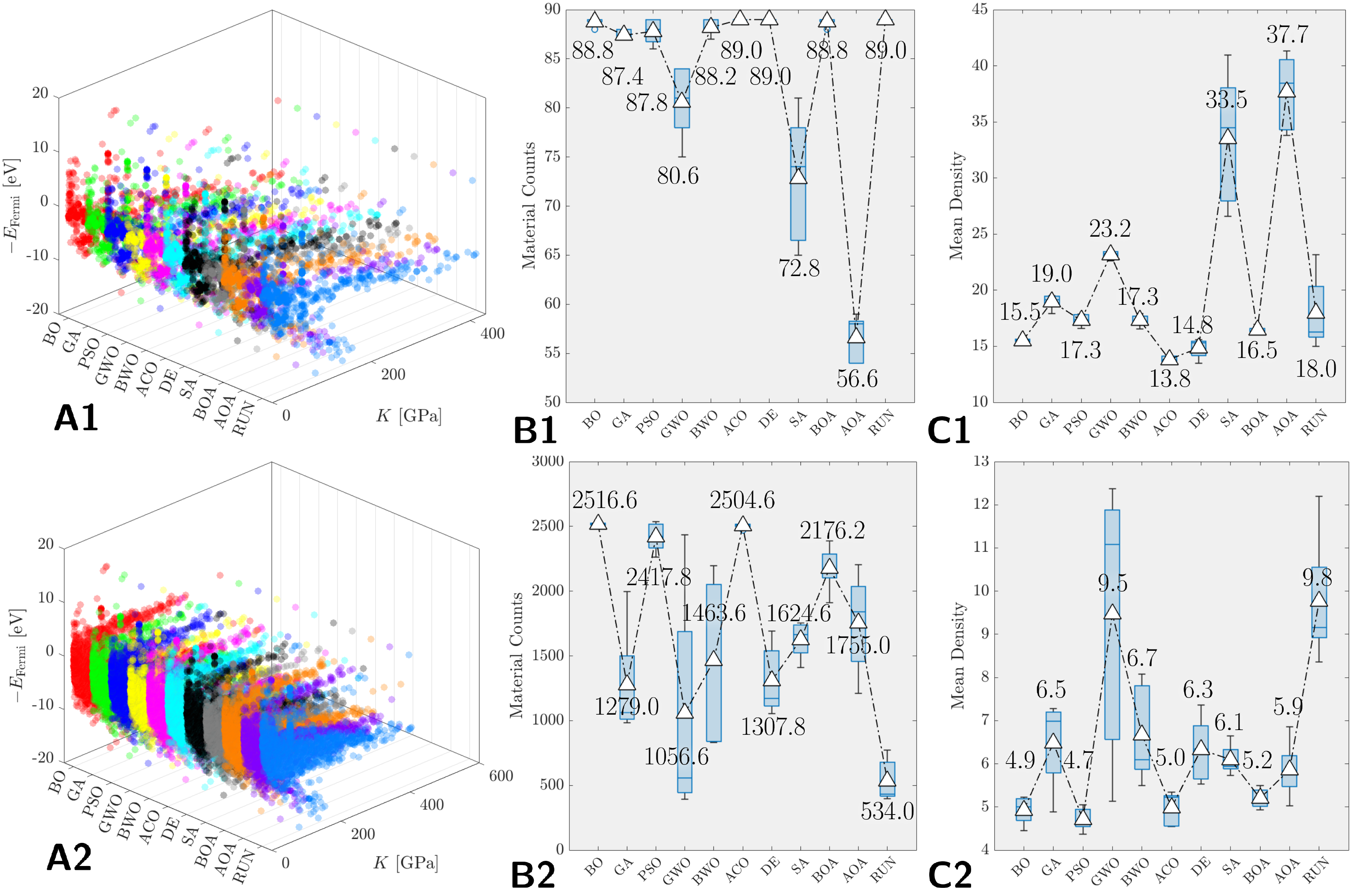}
    \caption{Objective properties exploration, objective value evaluation, and targeted design space count for benchmarking different optimization methods for single and multi-element composition design. {\bf (A1)} Overall design space benchmarking for the 11 different optimization schemes used, marked in different colors for the single-element composition design (For the projections in 2D, please refer to Figure \ref{sfig5}). {\bf (B1)} The distribution (for the 5 repeated experiments) and mean values of the overall material counts during the evaluation processes. {\bf (C1)} The distribution (for the 5 repeated experiments) and mean values of the scanned material evaluation densities (see Section \ref{sec_method} \& {\em Density Scan} in ESI for the calculation of ``Mean Density'') during the evaluation processes. (Related details can be found in the analysis of Figure \ref{sfig6}). {\bf (A2)} Overall design space benchmarking for the multi-element composition design (refer to Figure \ref{sfig5} for details). {\bf (B2)} The distribution and mean values of the overall material counts. {\bf (C2)} The distribution and mean values of the scanned material evaluation densities. The white triangular dots are the mean values for the 11 evaluations in both \textbf{(B)} \& \textbf{(C)}. The corresponding numbers are marked on top of the bar plot for both \textbf{(B)} \& \textbf{(C)}. BO, GA, PSO, GWO, HIWOA, ACO, DE, SA, BOA, AOA, \& RUN denote Bayesian Optimization, Genetic Algorithm, Particle Swarm Optimization, Hybrid Grey Wolf Optimization, Hybrid-Improved Whale Optimization Algorithm, Ant Colony Optimization, Differential Evolution, Simulated Annealing, Bees Optimization, Arithmetic Optimization, and Runge Kutta Optimization, respectively.}
    \label{fig3}
\end{figure}

Figure \ref{fig3} shows the benchmarks of the 11 different inverse optimization algorithms (IOAs) and their objective properties explorations (Column \textbf{A}), the distribution of the material evaluation numbers with the corresponding mean values (Column \textbf{B}), and the distributions of scanned density with overall mean values (Column \textbf{C}), for both the single (Row \textbf{1}) and multi-element (Row \textbf{2}) elemental composition design. In the objective distributions, different IOAs are marked in different colors, and the corresponding optimization methods are abbreviated. Here, the materials count represents the total number of unique elemental compositions scanned by the different IOAs. The mean densities of objective spaces are computed to estimate the repetitive evaluations of the same elemental compositions by computing the mean value of the density scan space (see Section \ref{sec_method} \& ESI). Both metrics help us understand the ``strategy'' of different algorithms when they scan the chemical design space to search for the ideal elemental compositions.

\begin{table}[htbp]
    \centering
    \begin{tabular}{c|c c c c}
        Optimizations & MC (Single-element) & MDS (Single-element) & MC (Multi-element) & MDS (Multi-element)\\\hline
        BO & 0.2& $6.5\times10^{-2}$& 104.3&  $10.1\times10^{-2}$\\
        GA & 0.3& $42.9\times10^{-2}$&179704.5&$101.7 \times10^{-2}$\\ 
        PSO & 1.7& $28.9\times10^{-2}$&12575.7&$7.1\times10^{-2}$\\
        GWO & 14.3& $13.3\times10^{-2}$ &771728.3&$1019.3\times10^{-2}$\\
        BWO  & 0.7& $26.0\times10^{-2}$ &402679.3&$131.2\times10^{-2}$\\
        ACO & 0&$9.7\times10^{-2}$ &427.3&$15.8\times10^{-2}$\\
        DE & 0&$79.6\times10^{-2}$ &73394.7&$56.8\times10^{-2}$\\
        SA & 45.2& $3576.7\times10^{-2}$&19843.8&$12.5\times10^{-2}$\\
        BOA & 0.2& $8.4\times10^{-2}$&29907.2&$5.2\times10^{-2}$\\
        AOA & 5.8& $1156.6\times10^{-2}$&150198.5&$44.1\times10^{-2}$\\
        RUN & 0& $1098.7\times10^{-2}$&27895.5&$218.3\times10^{-2}$\\
        \hline
    \end{tabular}
    \caption{The variances for the five repeated experiments of different optimization algorithms for single- and multi-element composition design, expressed with single-digit precision. Note that MC denotes ``Materials Count'', and MDS stands for ``Mean Density Scan''. The four columns correspond to Figure \ref{fig3} \textbf{B} \& \textbf{C}. For better presentation, we rescale the variances of MDS 100 times larger (indicated in the form $\times10^{-2}$).}
    \label{tab_1}
\end{table}

There are five major findings from Figure \ref{fig3}. (1) Comparing \textbf{A1} and \textbf{A2}, the scanned objective spaces are smaller for GWO compared with other optimization methods for both the single and multi-element composition design cases. For the single-element composition design case: (2) the evaluated elemental composition numbers are similar for most methods, whereas SA and AOA are observably smaller. Both GWO and SA display relatively high variances compared with other methods. (3) The mean scanned densities are relatively higher for SA and AOA methods, with both displaying higher variances. Both SA and AOA possess higher variances. For the multi-element composition design cases: (4) RUN has the lowest evaluated elemental composition numbers. GWO and BWO have the highest variances. (5) For the mean scanned densities for multi-element compositions, GWO and RUN have the highest evaluated elemental composition numbers and variances.


\begin{table}[htbp]
    \centering
    \begin{tabular}{c|c c c c}
        Optimizations & MO (Single-element)& Vr (Single-element)& MO (Multi-element)& Vr (Multi-element)\\\hline
BO & 91.9& $58.7\times10^{-2}$&84.5&$57.5\times10^{-2}$\\
GA & 97.2& $1097.1\times10^{-2}$&91.0&$3726.5\times10^{-2}$\\
PSO &93.0 &$90.7\times10^{-2}$ &84.0&$8.8\times10^{-2}$\\
GWO & 96.8& $580.6\times10^{-2}$&89.8&$3816.3\times10^{-2}$\\
BWO& 94.9& $838.0\times10^{-2}$& 91.0&$3719.7\times10^{-2}$\\
ACO &92.5 &$148.3\times10^{-2}$ &84.0&$307.5\times10^{-2}$\\
DE &94.2 & $557.5\times10^{-2}$&89.2&$593.5\times10^{-2}$\\
SA &97.6 &$81.5\times10^{-2}$ &82.6&$145.8\times10^{-2}$\\
BOA & 93.9& $404.5\times10^{-2}$&87.3&$314.1\times10^{-2}$\\
AOA & 94.7&$840.2\times10^{-2}$ &83.6&$523.2\times10^{-2}$\\
RUN & 96.7&$429.3\times10^{-2}$ &92.8&$166.6\times10^{-2}$\\\hline
    \end{tabular}
    \caption{The mean objective values and variances for the five repeated experiments of different optimization algorithms for single- and multi-element composition design, expressed with single-digit precision. Note that MO denotes ``Mean Objective'', and Vr stands for ``Variances''. The four columns correspond to Figure \ref{fig4} \textbf{A}. For better presentation, we rescale the variances of MDS 100 times larger (indicated in the form $\times10^{-2}$).}
    \label{tab_2}
\end{table}

\begin{figure}[htbp]
    \centering
    \includegraphics[scale=0.37]{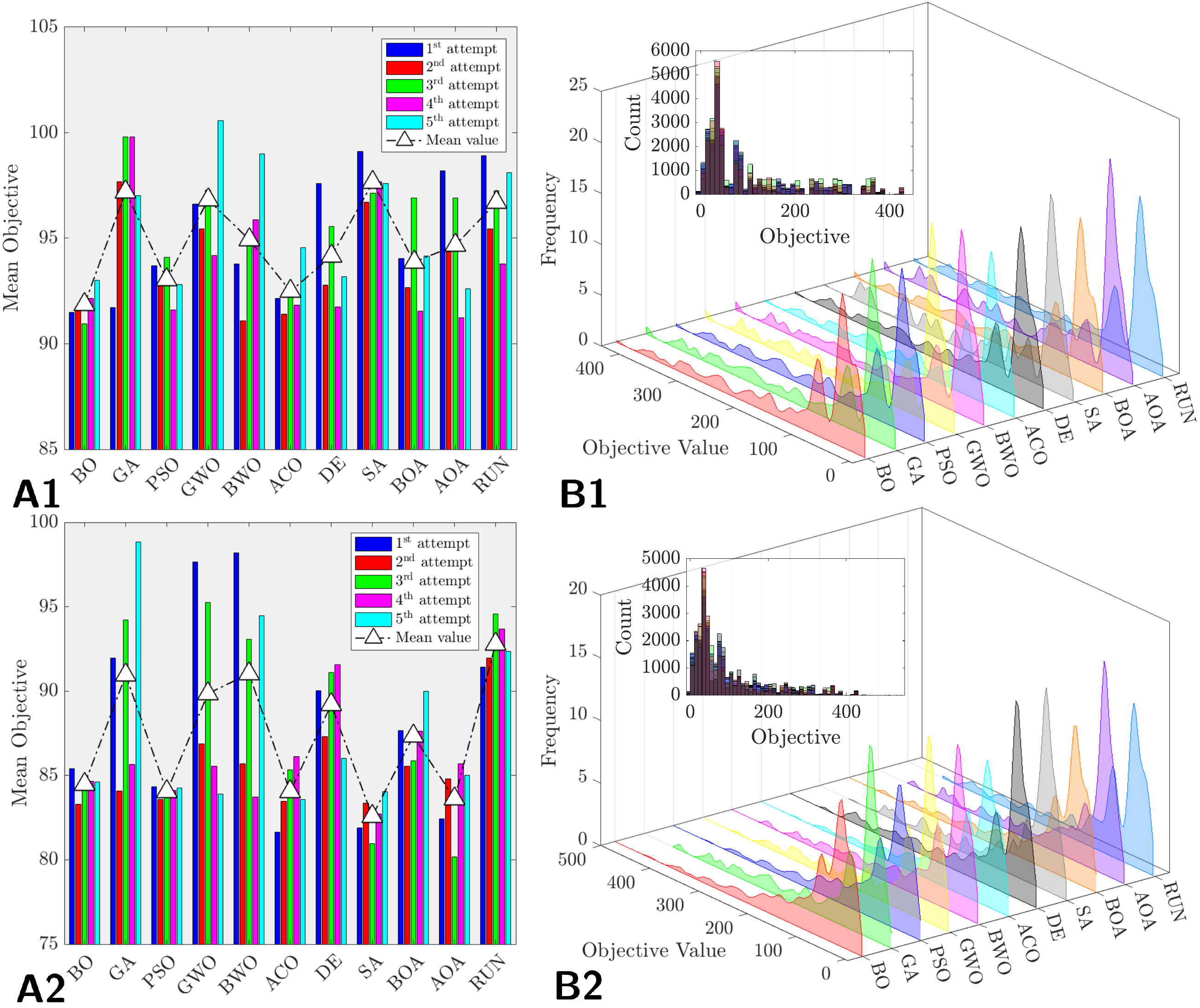}
    \caption{The statistical evaluations of the optimization processes. \textbf{(A1)} The distribution of the mean objective values (from 5 attempts) for the 11 evaluated optimization algorithms for single-element composition cases. Different colors mark different attempts, as indicated in the legend in the top-right corner of the figure. The white triangular dots stand for the mean values of the five attempts. The ``Mean Objectives'' are calculated as $\rm\frac{\sum(objectives)}{attempts}$ per optimization attempt (under the same random seed). \textbf{(B1)} The frequency distributions (i.e. repeated evaluations) using kernel fits of different objective values for the 11 different optimization algorithms (Details see Section \ref{sec_method} \& {\em Additional Results} in ESI). Note that the original data distribution is shown in the top left corner of the subfigure. \textbf{(A2)} The distribution of the mean objective values for the 11 evaluated optimization algorithms for multi-element composition cases, where the white triangular dots are the mean values. \textbf{(B2)} The frequency distributions of different objective values for the 11 optimization algorithms. }
    \label{fig4}
\end{figure}


From Figure \ref{fig4}, we can draw five key observations. (1) When comparing \textbf{A1} and \textbf{A2}, the mean objective distributions differ for the single-element and multi-element composition design cases. (2) For single-element design, GA, SA, and RUN display high mean objective values. SA and RUN also exhibit relatively low variances. GWO shows relatively high mean objectives but with higher variances, while BO demonstrates a stable performance with a low mean objective value and low variance. (3) DE and AOA exhibit repetitive scans in the same composition region, indicating a tendency to get "trapped" in a local space. This aligns with their relatively lower mean objectives in \textbf{A1}. Moving on to multi-element design, RUN achieves the highest mean objective values with low variances. GA, GWO, and BWO show higher variances. (4) AOA and SA again exhibit repeated scanning of the same elemental compositions, suggesting a tendency for optimization methods to get trapped in local composition spaces. (5) When considering both single and multi-element cases, different optimization algorithms demonstrate similar elemental composition search strategies, as evidenced by similar frequency distributions.

To further quantitatively support our observations on the uncertainties of different optimizations for repeated elemental composition design experiments in Figures \ref{fig3} \& \ref{fig4}, Tables \ref{tab_1} \& \ref{tab_2} show the variances among different experiments per different optimizations. Both Table \ref{tab_1} \& Figure \ref{fig3} show that the materials count variances are very different for single- and multi-element compositions, due to the very limited single-element compositions existing in the chemistry table. For multi-element composition design, BO has the lowest variance and GWO has the highest variance of materials count, with GWO's variance approximately 7398 times higher than BO. Other than that, GWO's variance is approximately 27 times higher than that of RUN of materials count. GA and AOA have approximately the same level of uncertainty for materials count. The relative magnitude of the variances of the materials count and the mean density scan does not follow the exact same trend. Yet for multi-element composition design, GWO still exhibits the highest variance, with 366.93\% higher than that of BO and 9992.08\% higher than that of BO. Generally, it can still be deduced from Table \ref{tab_1} that, for both single- and multi-element composition design, nature-inspired algorithms (i.e., GA, GWO, BWO) display relatively higher variances, indicating higher uncertainties, which will be further discussed in Section \ref{section_discussion}. Table \ref{tab_2} further confirms the observation that nature-inspired algorithms present higher uncertainties. For both single- and multi-element composition designs, GA, GWO, \& BWO show evidently higher variances of the mean objectives than that of other optimizations. For single-element composition design, GA presents 155.56\% and 1093.8\% relatively higher variances, GWO presents 35.24\% and 531.77\% relatively higher variances, and BWO presents 95.20\% and 811.86\% relatively higher variances than that of RUN and BO, respectively. For multi-element composition design, GA presents 2136.79\% and 6380.87\% relatively higher variances, GWO presents 2190.70\% and 6537.04\% relatively higher variances, and BWO presents 2132.71\% and 6369.04\% relatively higher variances than that of RUN and BO, respectively. 

\begin{figure}[htbp]
    \centering
    \includegraphics[scale=0.32]{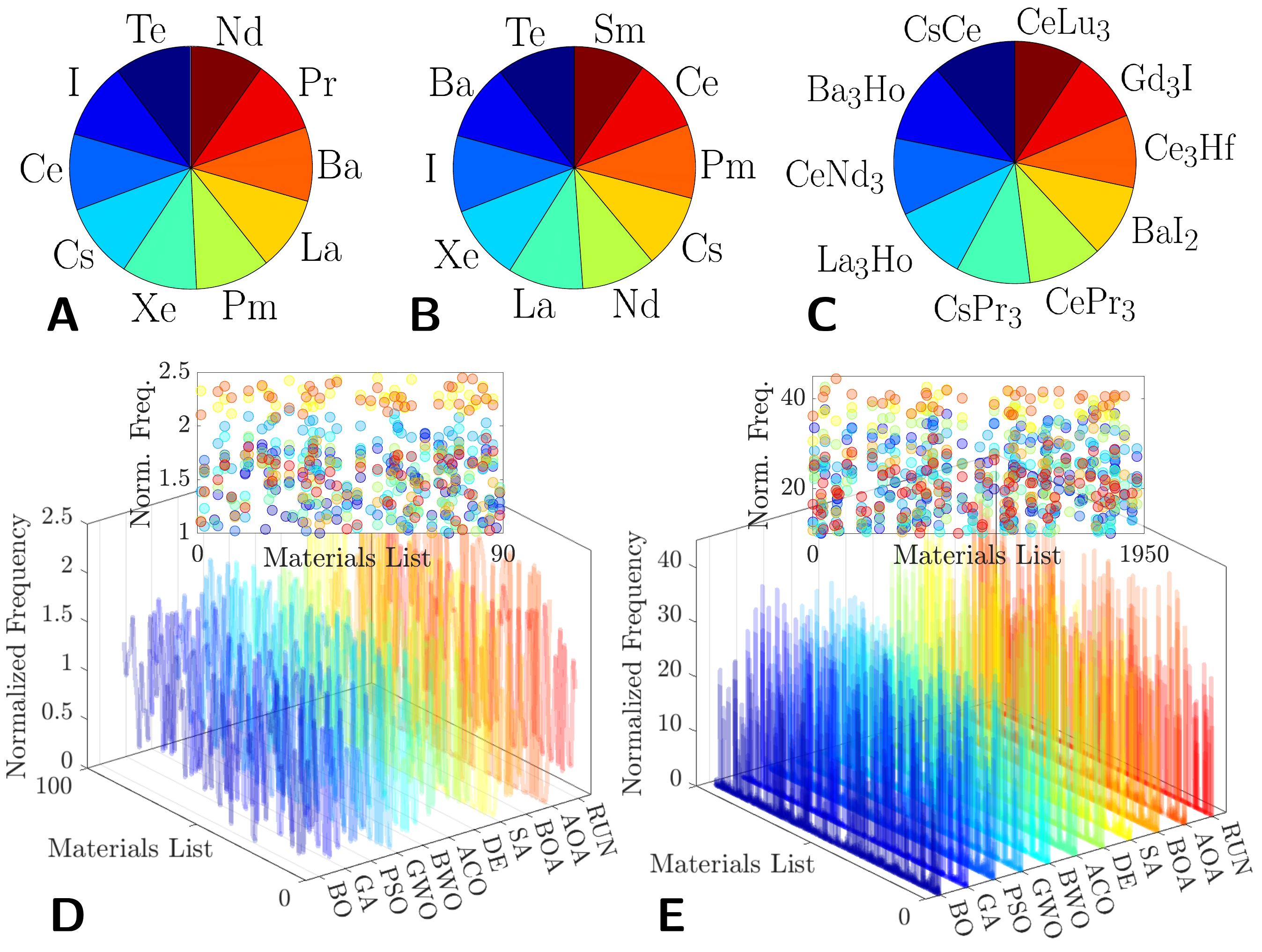}
    \caption{Extracted elemental composition evaluations for the optimization benchmarking. {\bf (A)} The top ten elemental compositions extracted from the single-element design case. {\bf (B)} The top ten elemental compositions extracted from the multi-element composition design cases. {\bf (C)} The top ten chemical compounds extracted from the multi-element composition design case. \textbf{(D)} The normalized frequency (see {\em Additional Results} in ESI for extended analysis for frequency evaluations of optimizations) for different extracted elemental compositions under the single-element composition design cases. The top subfigure dots stand for the projection of the normalized frequency in the range of $[1,2.5]$. \textbf{(E)} The normalized frequency under the multi-element composition design cases. The top subfigure dots stand for the projection of the normalized frequency in the range of $[10,45]$. }
    \label{fig5}
\end{figure}

Figure \ref{fig5} visualizes the material distribution generated by all the optimization algorithms, the changes in objective values during material evaluations benchmarked against silicon, and the reconstruction of the design space averaged across the 11 different optimizations. In Figure \ref{fig5} \textbf{A}, the crystal phases of Xe and Kr are repeatedly scanned by multiple algorithms, despite their instability under room temperature. This raises an intriguing question for computational optimization: How can complex factors such as cost and industrial applications be incorporated into the optimization framework to enhance its realism for real-world deployment? This may be left for future explorations. AOA and GWO show an increasing trend in the number of evaluated elemental compositions in the target space for bulk modulus, while BO, GA, and GWO exhibit the same trend for Fermi energy. Figure \ref{fig5} \textbf{C} visually depicts the design space, conveying a clear message that it is highly nonconvex and presents significant challenges in searching for the optimal point. Moreover, the non-continuous contours observed in the averaged design spaces based on evaluated elemental compositions further complicate the search for the global optimum. For a detailed discussion on the design spaces of all 11 optimization algorithms (and DRL), please refer to the ESI.

From Figure \ref{fig5}, we can derive five main observations. (1) Comparing \textbf{A} and \textbf{B}, we observe that the top evaluated elemental compositions are similar. (2) The majority of the top chemical compounds are binary alloys, consisting of elements present in the top single-element compositions such as Cs, Ce, Ba, and others. (3) RUN demonstrates effective scanning strategies and consistently converges towards local optimal elemental compositions. (4) SA demonstrates similar search strategies as RUN, evidenced by the high normalized frequency along the materials list. (5) BO appears to struggle in identifying the "ideal" elemental compositions and tends to scan the same region repeatedly, indicating a potential trap in the local optimal area. {Based on Figures \ref{fig3}-\ref{fig5}, we have also added qualitative comparison and suggested potential applications in Table \ref{tab3-method_comparison}. Since we hope to design materials digitally with low uncertainty and targeted desired properties quantified through objectives, low variance, and high mean objectives can be viewed as ``advantages'' of the algorithm. ``Targeted materials selection'' is recommended due to repeated materials sampling and ``Large-scale exploration'' is recommended based on more materials candidates sampled with decent objective values.}

\begin{table}[]
    \centering
\begin{tabular}{c|c c c}
        Optimizations & Advantages & Disadvantages & Potential Use\\\hline
BO &  Low variances & Low objective & Large-scale materials exploration\\
GA & High objectives & High variances & Targeted materials selection\\
PSO & Low variances & Low objective & Large-scale materials exploration\\
GWO & High objective& High variances & Not recommended\\
BWO& High objective& High variances & Not recommended\\
ACO &Low variances & Low objective & Large-scale materials exploration\\
DE & Decent objective & Relatively high variances &Not recommended\\
SA & High selectivity & Problem-specific & Not recommended\\
BOA & Decent objective& Low selectivity &Not recommended\\
AOA & High selectivity&High variance & Targeted materials selection\\
RUN & High objective \& stable & Low selectivity & Materials exploration \& design \\\hline
    \end{tabular}
    \caption{Assessment and conclusion from the comparison results for optimization methodologies benchmarked. Note that the (dis)advantages concluded from the previous results may be subjective. Note that the "Low selectivity" for RUN indicates the variance is relatively low compared with DE and AOA.}
    \label{tab3-method_comparison}
\end{table}

Among all the metrics we evaluated, it can be deduced that RUN achieved superior optimization results compared with the others (Figures \ref{fig3} \& \ref{fig4}, Tables \ref{tab_1} \& \ref{tab_2}). This out-performance may be due to several reasons. (1) RUN is free of hyperparameters. The update process of RUN does not depend on fine-tuned hyperparameters, making it more robust especially for highly nondeterministic situations like elemental composition design. (2) The updating strategies of RUN are very different than that of other methods. The RUN optimization searches for the next set of evaluations by mimicking the update strategies of Runge Kutta's discretization of differential equations (see ESI), which are fundamentally different from other metaheuristic methods. (3) From (2), the enhanced solution quality (ESQ) mechanism is well suited for our formulated elemental composition discovery cases. In the authors' original proposal \cite{RUN_optimization}, the ESQ mechanism is employed to avoid the local optimal solutions and increase convergence speed. In our case, as the material evaluation numbers are fixed at 5000, the faster convergence induced by ESQ enables the robust performance of the RUN algorithm in our benchmark cases. Further comprehensive characteristics are discussed in Section \ref{section_discussion}.

\section{Discussion\label{section_discussion}}


Based on the results presented earlier, our findings raise several intriguing questions, for which we proffer the following discussions. Firstly, we observe variations in the performance of different optimization algorithms across the two elemental composition design scenarios, such as the contrasting results of material counts and mean density of RUN and AOA (Figure \ref{fig3} \textbf{B} \& \textbf{C}). One explanation is that algorithm effectiveness is problem-specific. Moreover, the optimization methods may also be sensitive to the dimensionality of the problem, such that changing from single-element to multi-element composition design induces the change of metrics evaluation shown in Figures \ref{fig3} \& \ref{fig4}. Additionally, in the case of single-element composition design, the limited number of available elemental compositions can impact the performance of certain algorithms, potentially influencing their search capabilities. An interesting observation pertains to the consistent success of the RUN algorithm in achieving high mean objective values in both problem scenarios, especially in the second case (Figure \ref{fig4} \textbf{A}). This achievement can be attributed to the fact that the sampling strategy employed by RUN is hyperparameter-free and adept at storing information, which proves advantageous in handling high-dimensional spaces. 

Another intriguing finding is that BO exhibits lower mean objectives accompanied by low variances in both cases. This trend becomes more apparent when considering Figures \ref{fig4} \textbf{B} \& \ref{fig5} \textbf{E}. A plausible explanation would be BO is sampling a wider range of diverse elemental compositions during the optimization process. This behavior can be attributed to the uncertainty-based sampling strategy employed by BO, which facilitates the exploration of a diverse design space and aids in constructing a more accurate surrogate model. AOA tends to repeatedly sample similar elemental compositions over an extensive number of evaluations (Figure \ref{fig4} \textbf{B}). This behavior suggests that AOA may become trapped within a specific region of the design space, necessitating further investigation to understand the underlying causes and identify potential remedies. Finally, we explore the major disparity in SA's performance between the two elemental composition discovery cases. A possible explanation for this discrepancy is that SA is sensitive to the dimension of the optimization problem, leading to its different performances tested from the designed metrics in Figures \ref{fig3} \& \ref{fig4}.
Each optimization algorithm demonstrates unique characteristics and strategies that contribute to its specific outcomes. The low variances of the objectives in Figure \ref{fig4} could be explained by BO's low dependence on hyperparameters. BO aims to approximate an accurate surrogate model to optimize the objective, resulting in small mean objective values. This strategy utilizes uncertainty-based sampling, which decreases the overall objective value. The low uncertainty and variance in the final results are attributed to the small number of hyperparameters and the large sampling base. Additionally, BO exhibits high material counts, as observed in Figure \ref{fig3} \textbf{B}. GA heavily relies on hyperparameters such as mutation, offspring generation, and randomization (details see ESI). Strong dependence on hyperparameters leads to high uncertainty in the mean objective values for both cases (Figure \ref{fig4} \textbf{A1} \& \textbf{A2}). Moreover, GA repeated samples of the same elemental compositions once ideal properties are detected, resulting in high sampling density compared with BO, PSO, and other related methods (Figure \ref{fig4} \textbf{B1} \& \textbf{B2}). PSO, on the other hand, exhibits lower uncertainty and variance as it has fewer hyperparameters, contributing to its more stable performance, as depicted in Figure \ref{fig4} \textbf{A1} \& \textbf{A2}.

GWO is similar to GA in terms of hyperparameter dependency, resulting in high uncertainty and sampling density focused on specific objective regions (Figure \ref{fig4}). Notably, GWO displays higher uncertainty even in material counts and mean density, which differentiates it from GA (Figure \ref{fig3}). Similarly, BWO exhibits high uncertainty in mean objective values (Figure \ref{fig4} \textbf{A}), which may also be attributed to its dependence on hyperparameters and very similar searching strategies of mimicking animal dynamical behavior. However, BWO displays more material counts and less sampling density scan (Figure \ref{fig3} \textbf{B} \& \textbf{C}), potentially attributed to the setting of the encircling prey parameter, which enhances sampling stability. ACO employs continuous probability-based sampling, resulting in relatively less uncertainty in the mean objective values (Figure \ref{fig4} \textbf{A}). ACO has similar mean objective values as BO and PSO while exhibiting lower sampling density scans (Figure \ref{fig3} \textbf{C1}), indicating that ACO avoids oversampling the same regime for the single-element composition design case. DE algorithmically operates as a variant of GA, yielding comparable mean objective results (Figure \ref{fig4} \textbf{A}). With fewer hyperparameters compared to GA, DE exhibits lower uncertainty (Figure \ref{fig4} \textbf{A}). The evolutionary strategies of DE demonstrate similar material counts to GA, as seen in Figure \ref{fig3}, and repeated sampling for favorable materials, as depicted in Figure \ref{fig5}. SA's performance is observed to be more problem-specific, as both the material counts and density scan vary per different design cases (Figure \ref{fig3} \textbf{B} \& \textbf{C} and Figure \ref{fig4} \textbf{A}). SA exhibits low uncertainty for the mean objective values (Figure \ref{fig4} \textbf{A}) due to fewer hyperparameters and the implementation of a Markov chain for the acceptance-rejection sampling criteria (details see ESI). BOA shares similarities with GWO and BWO in terms of high uncertainty in the mean objectives (Figure \ref{fig4} \textbf{A}) and low material counts for multi-element composition design (Figure \ref{fig3} \textbf{B2}), as they use particle-based, animal behavior-inspired search criteria (details see ESI). However, BOA exhibits relatively lower variance (Figure \ref{fig4} \textbf{A}), indicating lower dependence on, and having fewer, hyperparameters.

AOA uses a distinct searching strategy based on mathematical symbolized operations. AOA exhibits high uncertainty with relatively low mean objectives (Figure \ref{fig4} \textbf{A}). AOA tends to oversample the local regime, leading to low material counts (Figure \ref{fig4} \textbf{B}). This behavior can be attributed to a large number of hyperparameters and the overdependence on certain factors such as {\tt MOA} and {\tt MOP} (details see {\em Arithmetic Optimization Algorithm} in ESI), resulting in a scaling factor that underestimates previous locations. Finally, RUN achieves high objectives with low uncertainty and avoids oversampling in specific objective regions (Figure \ref{fig4}). The material count is sensitive to dimensionality, yet all dimensions yield successful search results, as judged by the high overall mean objectives (Figure \ref{fig4} \textbf{A}). RUN is characterized by being hyperparameter-free, incorporating numerous intermediate hyperparameters to increase the dimensionality of the iteration scheme, and relying on previous solutions for updates based on search history. {Following our previous discussion on RUN's ESQ search mechanism and being hyperparameter-free, we would like to examine why RUN had outperformed other algorithms by having high objective values and low uncertainties. RUN employs an update scheme that has four update parameters\footnote{$k_1$, $k_2$, $k_3$, $k_4$ in Ref. \cite{RUN_optimization}}, each containing randomization coefficients\footnote{$rand_1$ \& $rand_2$ in Ref. \cite{RUN_optimization}} and are connected. This randomization and ``parameter-connection'' could help explain the low uncertainty observed in our numerical experiments. For the high objective, RUN's ability to avoid local optima as reported by Ahmadianfar et al.\cite{RUN_optimization} could be a possible explanation, in which they contend that the adaptive mechanism employed to update the parameter and the ESQ are the main mechanisms that assure a good transition from exploration to exploitation.}


\section{Conclusion}
In this paper, we propose a framework that can incorporate most state-of-the-art optimization algorithms to perform multi-elemental composition discovery and design for crystals. Our framework parameterizes the input elemental graph structure to tailor the corresponding bulk modulus $K$ and Fermi energy $E_{\rm Fermi}$. The key goal is to conduct a comprehensive benchmark for many of the state-of-the-art optimization methods and provide further insights for future implementation of these algorithms for elemental composition design. We optimized both single and multi-element composition discovery for maximized $K$ and minimized $E_{\rm Fermi}$ by formulating a single objective optimization. We benchmarked 11 different inverse optimization algorithms by fixing the material evaluations. We found that GA, GWO, and BWO exhibit higher variances, and BO and RUN display generally lower variances for materials count, density scan (Table \ref{tab_1}), and mean objectives (Table \ref{tab_2}). We further conclude that nature-inspired algorithms contain more uncertainties in elemental design cases, which can be attributed to the dependency on hyperparameters. RUN exhibits higher mean objectives whereas BO displayed low mean objectives compared with other optimization methods. Combined with materials count and density scan, it is proposed that BO samples more elemental compositions aiming to approximate a more accurate surrogate of the design space, and hence have lower mean objectives, yet RUN will repeatedly sample the discovered optimal elemental compositions for the successful search strategy (Figures \ref{fig4} \textbf{B} and \ref{fig5} \textbf{D} \& \textbf{E}). We also proffer detailed discussions on the exhibited results that correspond to each optimization's characteristics.
 





\section{Acknowledgements}

JY acknowledges support from the US National Science Foundation under awards CMMI-2038057, ITE-2236190, and EFMA-2223785, and the Cornell faculty startup grant. The authors also acknowledge the computational resources provided by the NSF Advanced Cyberinfrastructure Coordination Ecosystem: Services \& Support (ACCESS) program under grant BIO210063 and the computational resources provided by the G2 cluster from Cornell University.

\section{Author contributions statement}

H.Z. and H.H. conceived the research. H.Z. designed and conducted the experiment(s). H.Z., H.H., and J.Y. analyzed the results. J.Y. acquired the funding and resources. All authors reviewed the manuscript. 

\section{Additional information}


\textbf{Accession codes}: The codes for the optimization benchmarking are available at \url{https://github.com/hanfengzhai/materials-discovery-benchmark/tree/master/inverse_opt}. 

\noindent\textbf{Competing interests}: The authors declared no competing interests.





\bibliography{sample}

\clearpage

\setcounter{equation}{0}
\setcounter{figure}{0}
\setcounter{table}{0}
\renewcommand{\figurename}{\textsc{Figure}}
\renewcommand{\thefigure}{S{\arabic{figure}}}
\renewcommand{\tablename}{\textsc{Table}}
\renewcommand{\thetable}{S\arabic{table}}
\def\theequation{S\arabic{equation}}

\vspace{50pt}

\begin{center}
    \textsf{\bfseries\LARGE Supplementary Information: Benchmarking Inverse Optimization Algorithms for Materials Design}
    
\vspace{10pt}
{\Large\sf Hanfeng Zhai$^{\star}$\footnote{Email: \tt hz253@cornell.edu}, Hongxia Hao$^\dagger$\footnote{Email: \tt hongxiahao@microsoft.com}, Jingjie Yeo$^\star$\footnote{Email: \tt jingjieyeo@cornell.edu}}\\
\vspace{10pt}
{\large\sf $\star$Cornell University, Ithaca, NY, USA}\\\vspace{5pt}
{\large\sf $\dagger$Microsoft Research AI4Science, Beijing, China}\\
    \vspace{25pt}

\end{center}

\begin{center}
    \textsf{\bfseries\Large Summary}
\end{center}
 {\sf This supplementary information contains the extension for our problem formulation for detailed explanations. We then provide further details on hyperparameters and basic derivation for the 12 optimization algorithms employed (the 11 employed optimizations for repeated experiments with deep reinforcement learning). We then proffer some additional results in supplementary to our proposition in the main article.}

\vspace{5pt}

\begin{center}
    \textsf{\bfseries\Large Additional Results}
\end{center}

\begin{figure}[ht]
    \centering
    \includegraphics[width=0.9\textwidth]{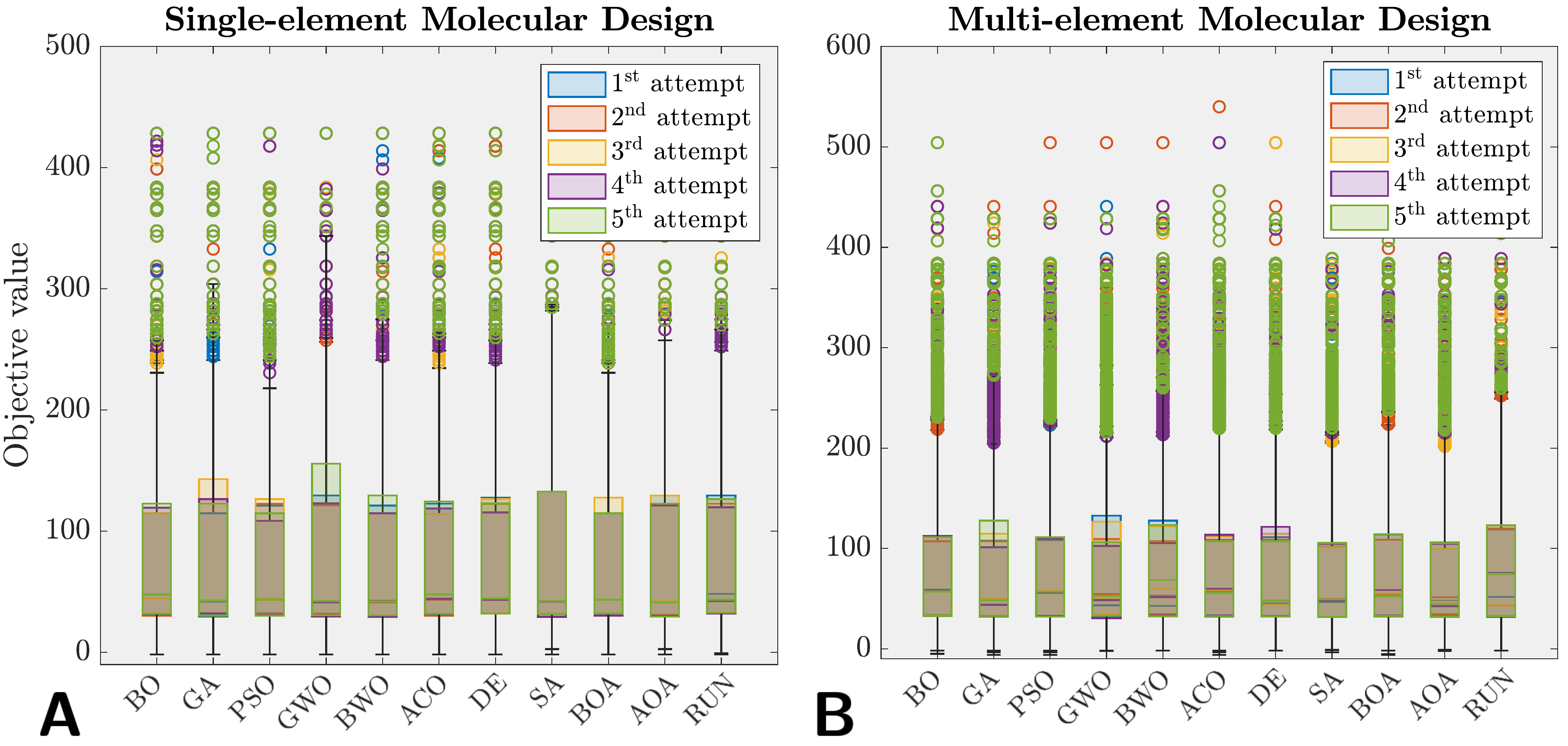}
    \caption{The objective value distribution over the five different experiments with different random seeds, in which the single-element composition design experiments are shown in \textbf{A} and multi-element case in \textbf{B}.}
    \label{sfig1}
\end{figure}

\begin{figure}[ht]
    \centering
    \includegraphics[width=0.9\textwidth]{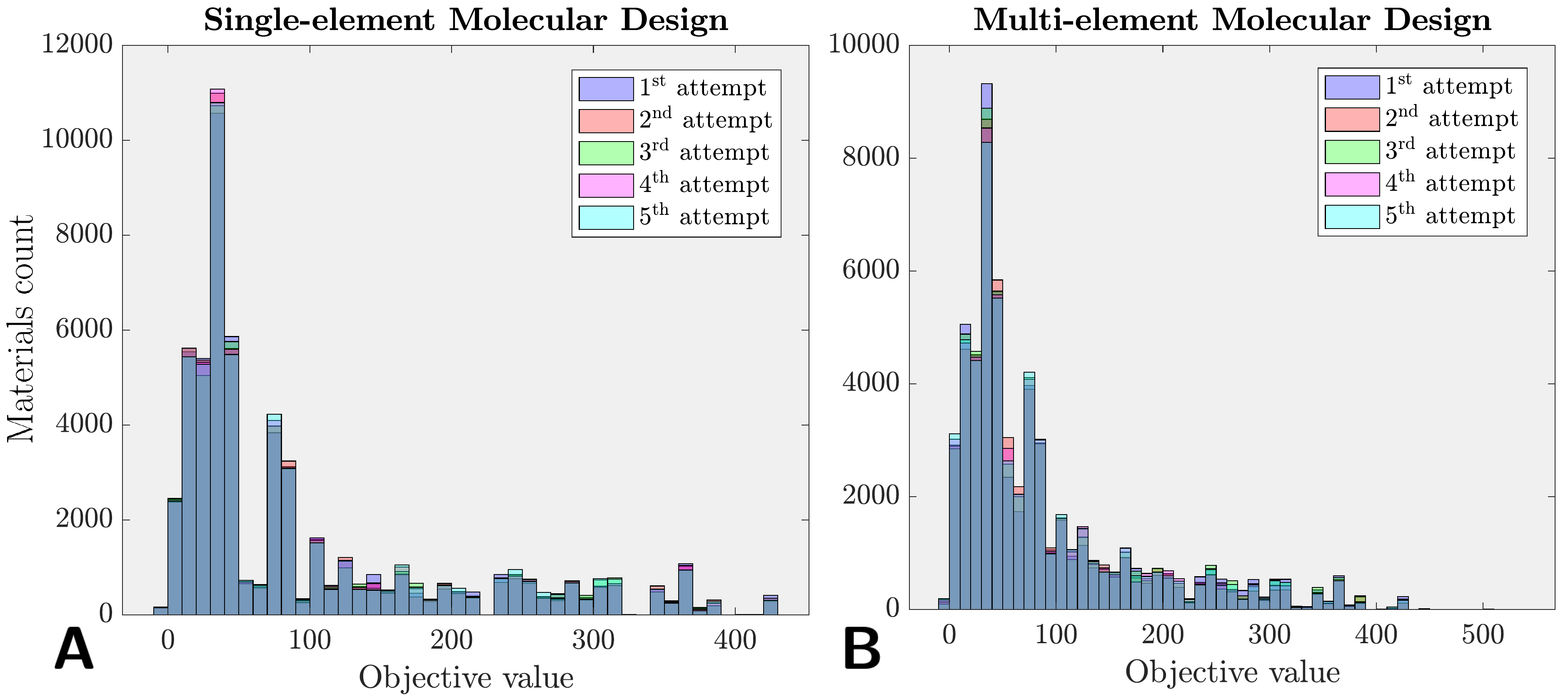}
    \caption{The distribution of the materials count over different objective values for the five different experiments, in which the single-element composition design experiments are shown in \textbf{A} and multi-element case in \textbf{B}.}
    \label{sfig2}
\end{figure}

Figure \ref{sfig1} shows the statistical distribution of the objective values over the five experiments with different random seeds. Figure \ref{sfig2} shows the statistical distribution of the materials count w.r.t. the objective values over the five experiments with different random seeds. It can be observed from both figures that the five repeated experiments generally match pretty well, indicating the robustness of the proposed framework in benchmarking optimization methods.

\begin{figure}[ht]
    \centering
    \includegraphics[width=0.9\textwidth]{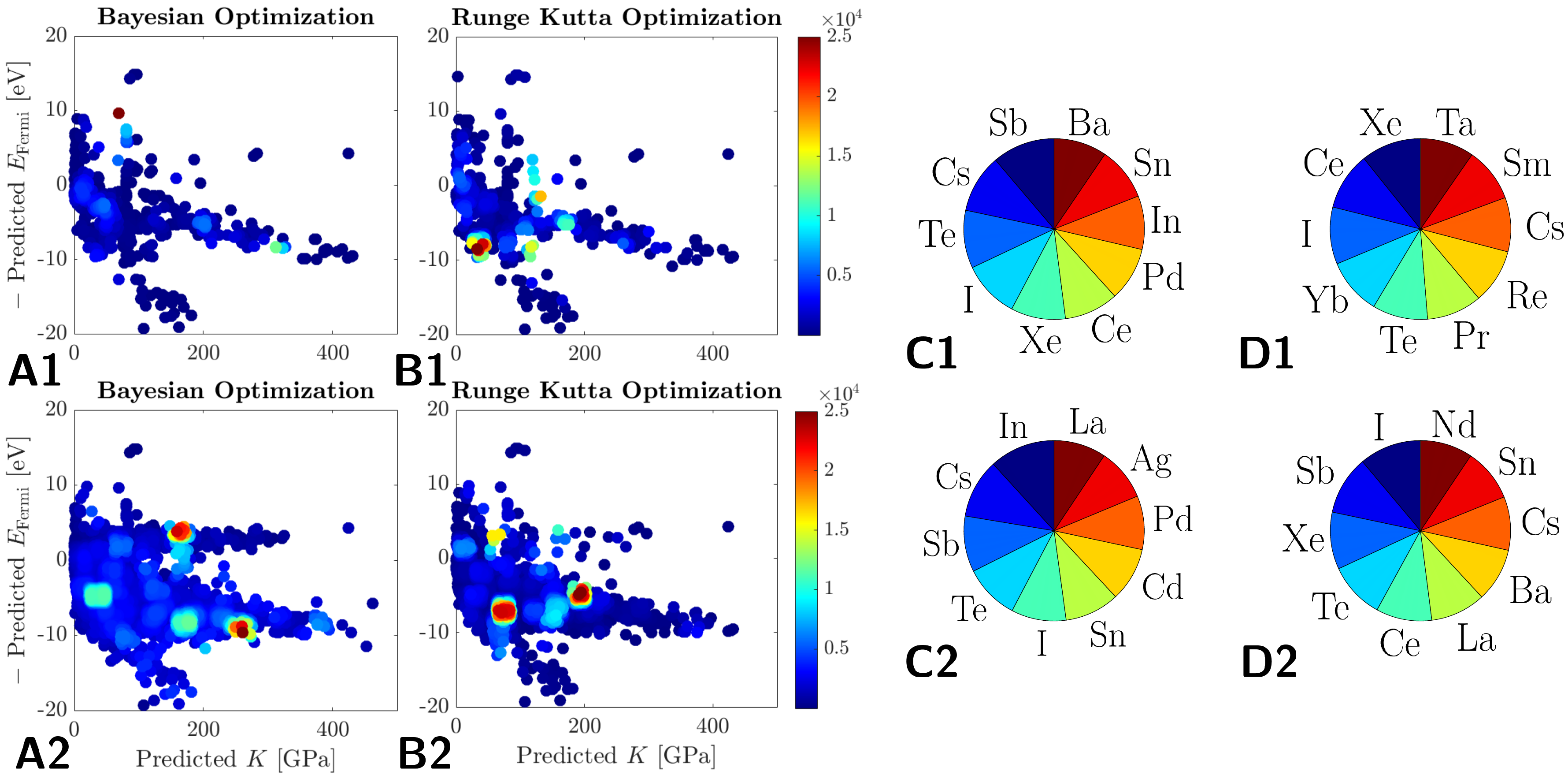}
    \caption{Some example numerical experiments showcasing results from Bayesian optimization and Runge Kutta optimization. Subfigures \textbf{A1} \& \textbf{B1} are the objective space evaluations for the single-element composition design for BO and RUN; Subfigures \textbf{A2} \& \textbf{B2} are the objective space evaluations for the multi-element composition design for BO and RUN. Subfigure \textbf{C1} \& \textbf{D1} are the top ten element compositions extracted from single-element composition discovery for the single-element composition design. Subfigure \textbf{C2} \& \textbf{D2} are the top ten element compositions extracted from single-element composition discovery for the multi-element composition design. }
    \label{sfig3}
\end{figure}

Figure \ref{sfig3} showcases the evaluation processes and the extracted element composition from both BO and RUN. The  evaluations in the objective spaces are marked with color gradients indicating the process, marked in the color bar on the right side of the figures (Figure \ref{sfig3} \textbf{A} \& \textbf{B}). The top ten element compositions are also visualized for comparison marked in color gradients (Figure \ref{sfig3} \textbf{C} \& \textbf{D}).

\begin{figure}[ht]
    \centering
    \includegraphics[width=0.55\textwidth]{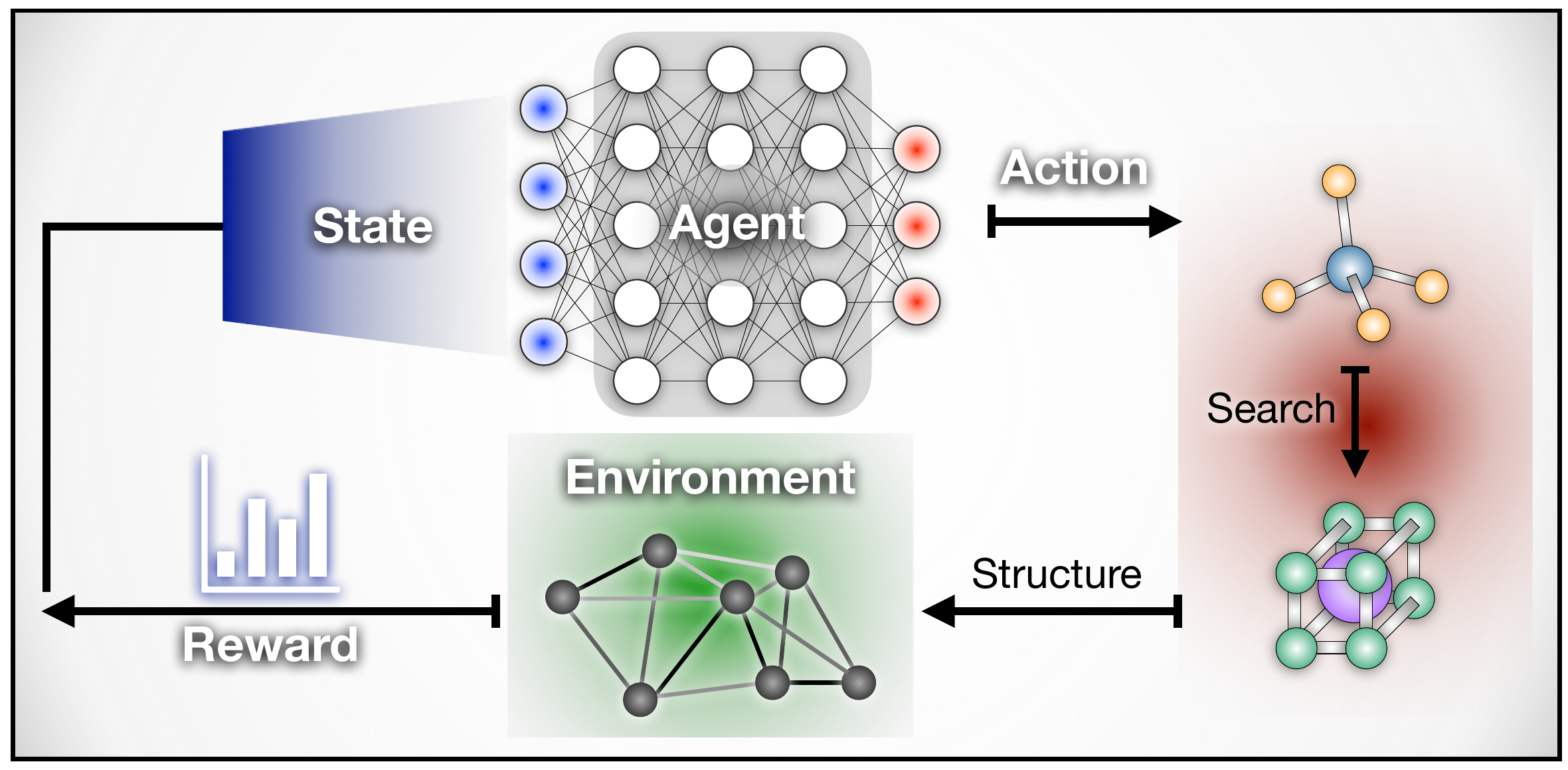}
    \caption{Schematic for materials discovery using deep reinforcement learning. The optimization is enabled by training a deep neural network as the "agent" to make optimal decisions, i.e., selecting the best materials. The action, and output of the network, screen the materials database and find the corresponding structure as input to the GNN as the environment. The environment then feeds back rewards (from predicted material properties) to the agent for making the best decisions. }
    \label{supp_fig_drl}
\end{figure}

\begin{figure}[ht]
    \centering
    \includegraphics[width=0.9\textwidth]{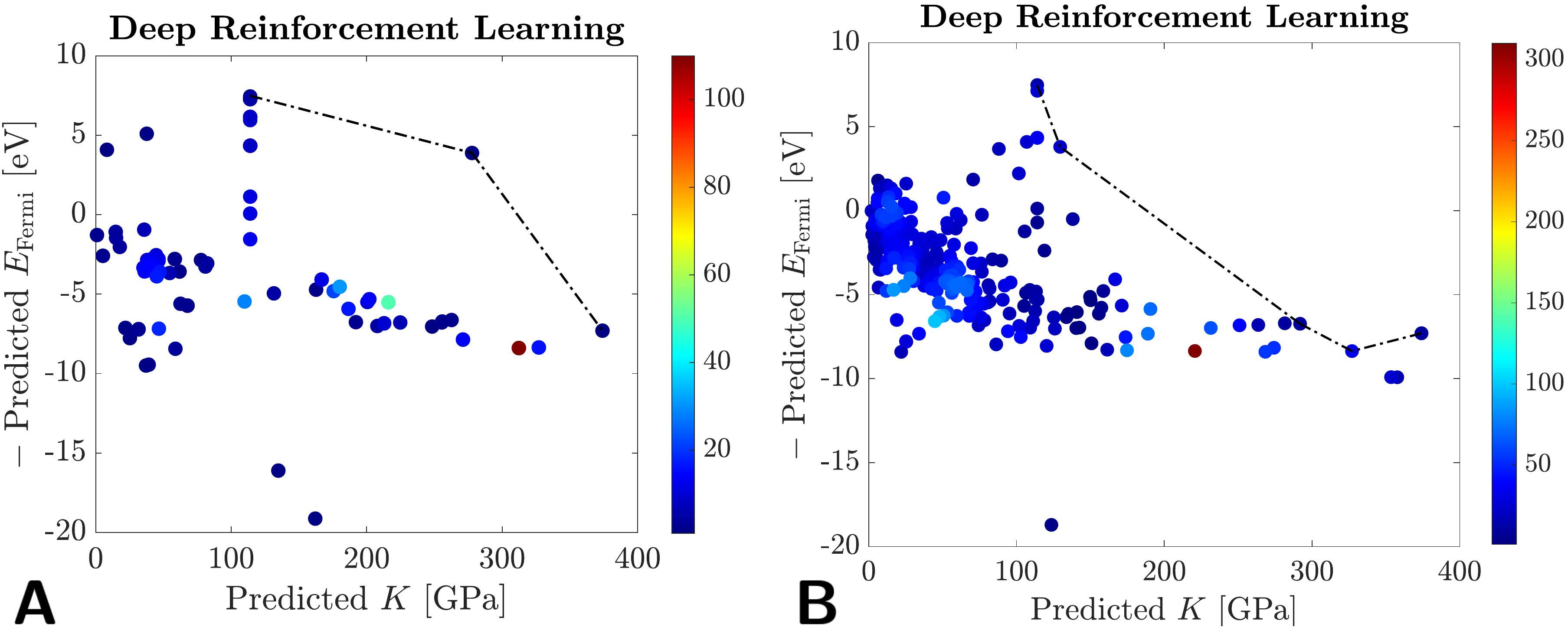}
    \caption{Example cases of materials design using the deep reinforcement learning methods. (\textbf{A}) Single-element composition design case with 110  evaluations. (\textbf{B}) Multi-element composition design case with 300  evaluations.}
    \label{sfig4}
\end{figure}

\begin{figure}[ht]
    \centering
    \includegraphics[width=0.9\textwidth]{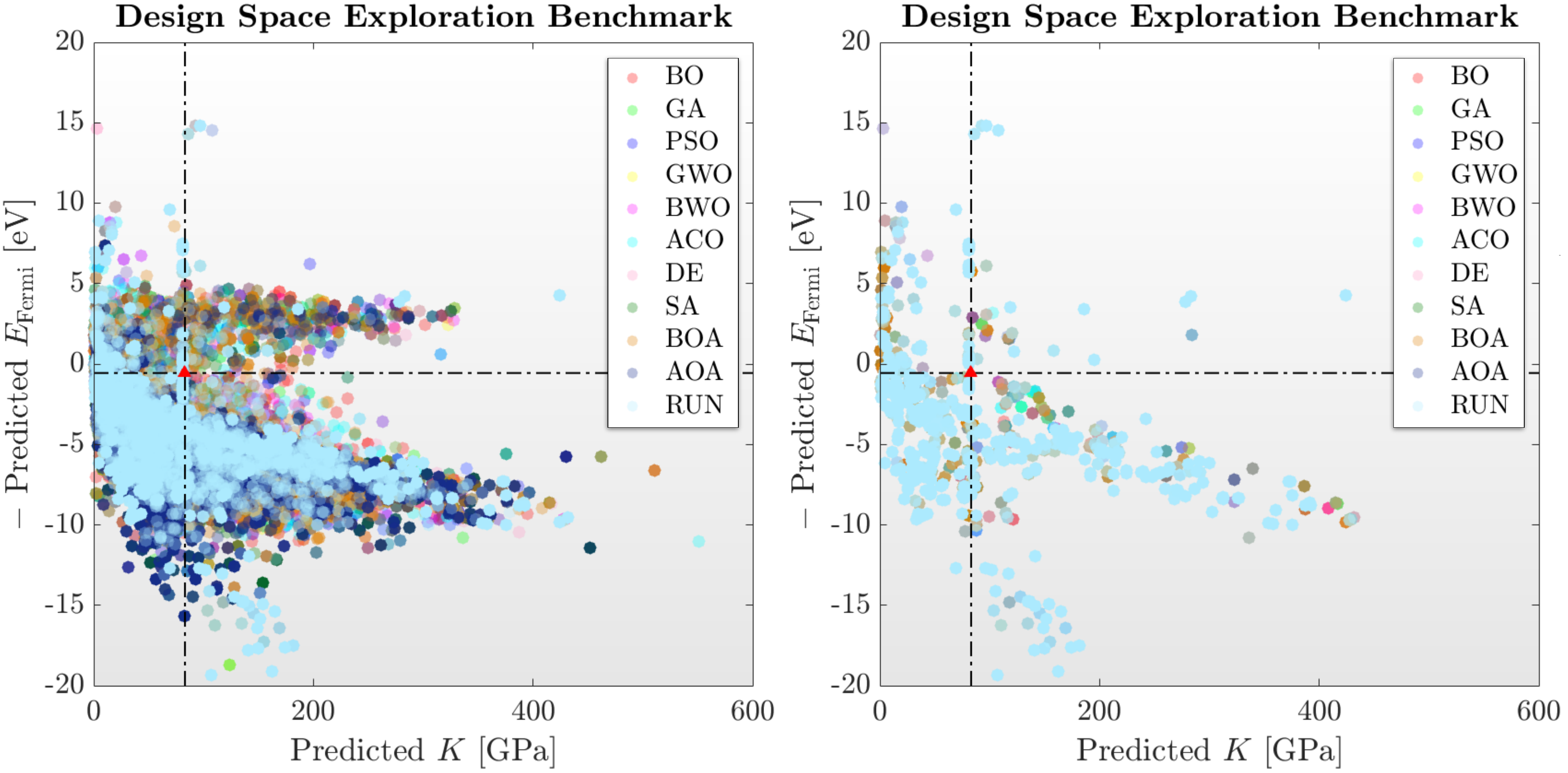}
    \caption{The projection in 2-dimensional objective spaces of different optimization algorithms corresponding to Figure \ref{fig3} \textbf{A}. The left subfigure is the multi-element composition design case and the right subfigure is the single-element composition design case. }
    \label{sfig5}
\end{figure}

Figure \ref{supp_fig_drl} shows the general schematic for implementing the deep reinforcement learning method in our proposed framework. The MEGNet material evaluations are treated as the environment and the agent (represented as a DNN) is trained to learn the action for selecting the best element composition from the materials project database. Detailed formulations are elaborated in the following section. Figure \ref{sfig4} showcases some preliminary results from using DRL to design materials, where the single-element composition design processes are shown in Figure \ref{sfig4} \textbf{A} and the multi-element composition design processes are shown in Figure \ref{sfig4} \textbf{B}. Correspondingly, the material evaluations indicated in Figure \ref{fig3} \textbf{A} are projected into 2-dimensional space and shown in Figure \ref{sfig5}.

\begin{figure}[ht]
    \centering
    \includegraphics[width=0.9\textwidth]{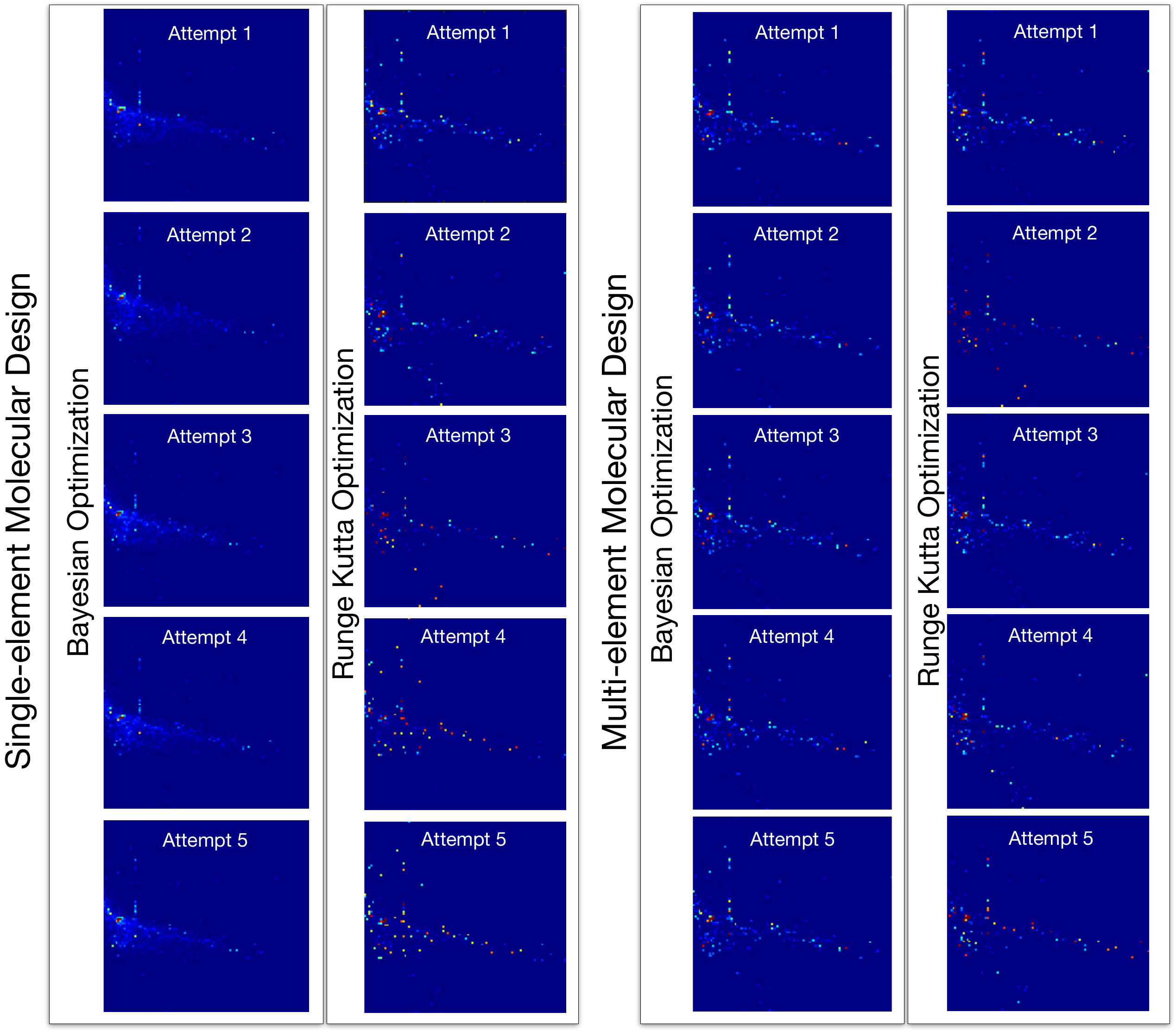}
    \caption{Example density scan map corresponding to the mean density scan evaluations in Figure \ref{fig3} \textbf{C} for BO and RUN. Note that the blue area corresponds to the zero  evaluation region. }
    \label{sfig6}
\end{figure}

To better visualize our efforts in estimating the design processes through the mean density scan methods, we present some example density scan evaluations using BO and RUN in Figure \ref{sfig6}. It can be seen that BO tends to evaluate a broad range of different element compositions with shallow gradients distributed among a material region, whereas RUN tends to repeatedly evaluate the detected ``optimal'' element composition multiple times, with sharp gradient points discretely distributed among the objectives space. This direct observation cross-validated our observations and proposed understandings of the searching strategies of BO and RUN in Sections \ref{section_results} \& \ref{section_discussion}.

\vspace{5pt}

\begin{center}
    \textsf{\bfseries\Large Extension on Problem Formulation}
\end{center}

Based on our problem formulation in Equation (\ref{eqn_opt_problem}), we can further expand the materials parametrization function $\Omega$. To explain, this is a highly nonconvex problem where the design variables are not related to the objective values at all. The ideal scenario, as described in the article, is to directly tune the graph structure representing the element composition to achieve direct modifications of its corresponding materials' properties. However, the difficulty that arises in this matter is it is knotty to represent the graph as the same dimensional input for perturbing the variables.

If assume the graph structures of the element composition can be written in the form of $\mathcal{G}({\bf u},{\bf e},{\bf v})$. The relation between the design variables and the objective values can be written as\begin{equation}
    \begin{aligned}
        K = MEGNet_K (\mathcal{G}({\bf u},{\bf e},{\bf v})) \xleftarrow[]{\mathcal{G}({\bf u},{\bf e},{\bf v}) = \Omega(n_{atom},\xi_n,\eta)}n_{atom},\xi_n,\eta\\E_{\rm Fermi} = MEGNet_{E_{\rm Fermi}} (\mathcal{G}({\bf u},{\bf e},{\bf v})) \xleftarrow[]{\mathcal{G}({\bf u},{\bf e},{\bf v}) = \Omega(n_{atom},\xi_n,\eta)}n_{atom},\xi_n,\eta
    \end{aligned}
\end{equation}where $MEGNet_K$ and $MEGNet_{E_{\rm Fermi}}$ represent the two pretrained models for predicting the properties of bulk modulus $K$ and Fermi energy $E_{\rm Fermi}$.
 In our formulation (Equation (\ref{eqn_opt_problem})), the graph is generated from the design variables via a parameterization function $\Omega$. This parameterization process relaxes the dimensional constraint of the graphs for different types of compounds. The function, $\Omega$, can be extended to\begin{equation}
     \begin{aligned}
     \mathcal{G}_{mat}(\mathbf{u},\mathbf{e},\mathbf{v}) = \Omega(n_{atom},\xi_n, \eta):&\\
         n_{atom} &= {\tt rand}_{\Theta_1} \in [n_{atom}^{low},n_{atom}^{high}],\\
         \xi_{n} &= {\tt rand}_{\Theta_2} \in [0,1],\\
         \eta &= {\tt rand}_{\Theta_3} \in [0,1],\\
         C_{mat} &= C_{mat}(n_{atom}, \xi_n),\\
         \mathbb{L}_{mat} &= \mathbb{L}_{mat}(C_{mat}) \in \left[MP\right],\\
         \mathcal{T}_{mat} &= \mathcal{T}_{mat}(\mathbb{L}_{mat}, \eta),\\
         \mathcal{G}_{mat}(\mathbf{u},\mathbf{e},\mathbf{v}) &= MP(\mathcal{T}_{mat})
     \end{aligned}
 \end{equation}where $\bf u,e,v$ are state, bond, and atom attributes, following the notation of Ref. \cite{megnet}. $n_{atom}^{low}$ and $n_{atom}^{high}$ are readjusted for the two design optimization scenarios. $C_{mat}$, $\mathbb{L}_{mat}$, and $\mathcal{T}_{mat}$ are different functions relating the design variables to the graph structure through hierarchical algorithmic steps. $[MP]$ simply stands for the material database stored in the {\em Materials Project}.

\section*{Estimation Metrics}

\subsection*{Density Scan}

We use density scans to estimate the repeated evaluations of the same material in the objectives space. First, the material evaluation in the objectives space (Figure \ref{fig3} \textbf{A} \& Figure \ref{sfig6}) is first projected on a $100\times 100$ grids as a density scan to describe the repeated evaluations of the same points. We then eliminate all the zero-value scan points and compute the mean values of the non-zero density scans. The equation describing the density scan can be written as\begin{equation}
    \bar{\rho} = \mathtt{mean}\left(\mathbb{N}(n_{atom},\xi_n,\eta)|\mathbb{N} \neq 0\right)
\end{equation}where $\mathbb{N}$ is the density scan, which is a function of the design variables $\mathbb{N} = \mathbb{N}(n_{atom},\xi_n,\eta)$.


\clearpage

\setcounter{equation}{0}
\setcounter{figure}{0}
\setcounter{table}{0}
\renewcommand{\figurename}{\textsc{Figure}}
\renewcommand{\thefigure}{S{\arabic{figure}}}
\renewcommand{\tablename}{\textsc{Table}}
\renewcommand{\thetable}{S\arabic{table}}
\def\theequation{S\arabic{equation}}
\begin{center}
    \textsf{\bfseries\Large Machine Learning Based Algorithms}
\end{center}
\section*{Deep Reinforcement Learning}
For approach (2), one trains a deep neural network (DNN) as the agent to learn the policy between the environment ($\Theta\rightarrow\mathcal{J}$) and action ($a$). Here, the environment would be the material evaluator, i.e., MEGNet. The action $a$ would be the change of the design variables $\Theta$. The process of agent tuning the design variables can be represented as \begin{equation}
    \Theta_{next} \xleftarrow{a} DNN (\mathcal{J}, r; \mathbf{p}_{DNN})
\end{equation}where $r$ is the rewards from the evaluated material properties, which the DNN receives as input. Similar to GPR, $\mathbf{p}_{DNN}$ is the hyperparameters involved in training the DNN. The agent receives the reward $r$ based on the properties of the evaluated material and takes the corresponding actions $a$ to tune the design variables $\Theta$.

Here, we employ the basic deep Q-learning. The update of the Q-learning schemes is solved from the Bellman's equation\cite{sutton2018reinforcement}:\begin{equation}
    Q^\pi(s,a) =\mathbb{E}\left[ r + \gamma\mathbb{E}\left[ Q^{\pi}(s',a')\right]\right]
\end{equation}where $Q^\pi$ is the optimal action-value function, $\gamma$ is the discount-rate parameter. $s'$ \& $a'$ represents the update for the action and states. The update for the agent's new action requires a {\em policy}, in which we employ the Boltzmann Q policy. 
In the DRL implementation, the optimization objective $\max\mathcal{J}$ is expected to be learned as policies via agent-environment interactions by feeding rewards. Here, we implement the basic Deep Q Network (DQN) by employing the Q-learning strategy. The agent numerically approaches the strategy for selecting the best $\Theta$ to either maximize or minimize $\mathcal{J}$.

In our DRL implementations, the agent is a deep neural network (DNN):\begin{equation}
        \begin{aligned}
            \left[a_1, a_2, a_3\right] = \left(L^5 \circ\sigma_{ ReLU} \circ L^4_{64} \circ\sigma_{ ReLU}\circ L^3_{64}\circ \sigma_{ ReLU}\circ L^3_{64} \circ \sigma_{ ReLU}\circ L^1_{32}\right) s,\\{\rm where}\ s \equiv \mathcal{J} = K - E_{\rm Fermi}
        \end{aligned}
    \end{equation}It can be seen that our agent has 3 hidden layers with 64 neurons per layer activated by the ReLU function. $s$ is the state, which we defined as the objective value. Here, $a_1$, $a_2$, $a_3$ correspond to changing the design variables, $n_{atom}$, $\xi_{n}$, $\eta$.
    The rewards are crafted based on the state as \begin{equation}
\begin{aligned}
    r = 
\begin{cases} 
     10, & s\geq 350 \\
     5, & 250 \leq s \leq 350 \\
     1, & 150 \leq s \leq 250 \\
     -5, & 50 \leq s \leq 150 \\
     -10, & s \leq 50
   \end{cases}
\end{aligned}
\end{equation}

The reward differences for different materials' properties intervals are relatively small. Our goal here is to pose a more challenging learning task for the agent to approximate the correct policy for maximizing the policy for design. The reward feeds into the total scores per episode to train the agent.

The standard reinforcement learning setup is considered, consisting of an agent interacting with an environment $E$ in discrete timesteps. The action-value function is used in many reinforcement learning algorithms. It describes the expected return after taking an action at in state st and thereafter following policy $\pi$: $Q^\pi(s, a) = \mathbb{E}[R|s, a]$. The optimization goal can be achieved via solving the Bellman equation, where the $Q^\pi$ values are updated recursively\cite{drl_control}:\begin{equation}
    Q^\pi(s,a) =\mathbb{E}\left[ r + \gamma\mathbb{E}\left[ Q^{\pi}(s',a')\right]\right]
\end{equation}

The Q-learning is achieved by minimizing the loss by considering the function approximators parameterized by $\mathbf{p}^Q$:\begin{equation}
    \mathcal{L}(\mathbf{p}^Q) = \mathbb{E}\left[\left(Q(s,a|\mathbf{p}^Q) - \mathcal{Y}\right)^2\right]
\end{equation}where $\mathcal{Y} = r(s,a) + \gamma Q(s',\mu(s')|\mathbf{p}^Q)$. The optimization is achieved via feedback-action loops through the rewards. Randomized \begin{tikzcd}
a \arrow[r, "MEGNet"] & s(a) \arrow[r, "DNN"] & r 
\arrow[ll, bend right=60, "DNN" description]
\end{tikzcd}
     Here, we employ the Boltzmann Q policy for the $Q^\pi$ values:
     \begin{equation}
         \pi(a|s) = \frac{e^{Q(s,a)/\tau}}{\sum_{a'} e^{Q(s,a')/\tau}}
     \end{equation}where $\pi(a|s)$ is the probability of taking action $a$ in state $s$; $Q(s,a)$ is the estimated Q-value of taking action $a$ in state $s$; $\tau$ is the temperature parameter that controls the exploration-exploitation trade-off. In our implementation, The algorithm undergoes an exploration phase consisting of 30 episodes, during which 5 elemental compositions are evaluated per episode. After this, the algorithm is trained for 60 steps, with 1 material evaluation per step. Finally, the algorithm is tested over 20 episodes, with 5 material evaluations per episode. This setup allows for the algorithm to learn and optimize material designs, with the testing phase providing a measure of its effectiveness.

\section*{Bayesian Optimization}

\subsection*{Gaussian Process Regression}


Gaussian process regression (GPR) is a Bayesian statistical approach to approximate and model function(s). Considering our optimization problem, if the function is denoted as $y = \hat{DS}(\mathbf{x, p})$, where $\hat{DS}$ is evaluated at a collection of different sets of points: $\mathbf{x}_1, \mathbf{x}_2, ..., \mathbf{x}_k \in \mathbb{R}^d$, we can obtain the vector $[f(\mathbf{x}_1), ..., f(\mathbf{x}_k)]$ to construct a surrogate model for the design parameters with the correlated objectives. The vector is randomly drawn from a prior probability distribution, where GPR takes this prior distribution to be a multivariate normal with a particular mean vector and covariance matrix. Here, the mean vector and covariance matrix are constructed by evaluating the mean function $\mu_0$ and the covariance function $\Sigma_0$ at each pair of points $x_i$, $x_j$. The resulting prior distribution on the vector $[f(x_1),..., f(x_k)]$ is represented in the form of a normal distribution to construct the surrogate model\begin{equation}
    \hat{DS}(\mathbf{x}_{1:k}) \sim \mathcal{N}\left(\mu_0 (\mathbf{x}_{1:k}), \Sigma_0 (\mathbf{x}_{1:k}, \mathbf{x}_{1:k}))\right)\label{surrogate}
\end{equation}
where $\mathcal{N}(\cdot)$ denotes the normal distribution. The collection of input points is represented in compact notation: $1:k$ represents the range of $1,2,..., k$.
The surrogate model $f(\mathbf{x})$ on $1:k$ is represented as a probability distribution given in Equation (\ref{surrogate}). To update the model with new observations, such as after inferring the value of $f(\mathbf{x})$ at a new point $\bf x$, we let $k = l+1$ and $\mathbf{x}_k = \mathbf{x}$. The conditional distribution of $f(\mathbf{x})$ given observations $\mathbf{x}_{1:l}$ using Bayes' rule is
\begin{equation}
    \begin{aligned}
   \hat{DS}(\mathbf{x})| \hat{DS}(\mathbf{x}_{1:l}) &\sim \mathcal{N}(\mu_l (\mathbf{x}), \sigma_l^2 (\mathbf{x}))\\
    \mu_l (\mathbf{x}) &= \Sigma_0 (\mathbf{x}, \mathbf{x}_{1:l}) \Sigma_0 (\mathbf{x}_{1:l},\mathbf{x}_{1:l})^{-1} \left(f(\mathbf{x}_{1:l}) - \mu_0 (\mathbf{x}_{1:l})+\mu_0(\mathbf{x}) \right)\\
    \sigma_l^2 &= \Sigma_0 (\mathbf{x}, \mathbf{x}) - \Sigma_0 (\mathbf{x}, \mathbf{x}_{1:l})\Sigma_0 (\mathbf{x}_{1:l}, \mathbf{x}_{1:l})^{-1} \Sigma_0 (\mathbf{x}_{1:l}, \mathbf{x})
    \end{aligned}
\end{equation}where the posterior mean $\mu_l(\mathbf{x})$ is a weighted average between the prior $\mu_0(\mathbf{x})$ and the estimation from $f(\mathbf{x}_{1:l})$, where the weight applied depends on the kernel used. We used the Matérn function in the implementation.


\subsection*{Acquisition Function}

Given the training data $[y_k,\ \mathbf{x}_k]$, Equation (\ref{surrogate}) gives us the prior distribution $y_l \sim \mathcal{N}(\mu_0, \Sigma_0)$ as the surrogate. This prior and the given dataset induce a posterior: the acquisition function denoted as $\mathcal{A}: \mathcal{X}\xrightarrow{} \mathbb{R}^+$, determines the point in $\mathcal{X}$ to be evaluated through the proxy optimization $\mathbf{x}_{ best} = \argmax_{ x}\mathcal{A}(\mathbf{x})$. The acquisition function depends on the previous observations, which can be represented as $\mathcal{A} = \mathcal{A}(\mathbf{x}; (\mathbf{x}_l, y_l), \theta)$. Taking our previous notation, the new observation is probed through the acquisition\cite{bo_biofilm} \begin{equation}
    \mathbf{x}_{k}=\mathbf{x}_{l+1}= \argmax_{x\in \frac{\mathcal{X}}{\mathcal{X}_l}} \mathcal{A}\left( \mathbf{x};(\mathbf{x}_l, y_l), \theta_m\right)
\end{equation}where the input space contains the evaluation of design variables at $n$ points: $\mathcal{X}_l : = (\mathbf{x}_1, \mathbf{x}_2, ..., \mathbf{x}_l)$. In our case, $\mathcal{X}$ is acquired through running $n$ numbers of material evaluations. We pick the GP Upper Confidence Bound (GP-UCB) as the acquisition function, exploiting the lower confidence bounds (in the case of minimizing the objective function) to construct the acquisition and minimize the regret. GP-UCB takes the form\begin{equation}
    \mathcal{A}\left({\bf x}; (\mathbf{x}_l, y_l), \theta_m\right):= \mu_l \left(\mathbf{x}; (\mathbf{x}_l, y_l), \theta_m\right) + \kappa \sigma\left(\mathbf{x}; (\mathbf{x}_l, y_l), \theta_m\right)\label{acquisition_function_equation}
\end{equation} where $\kappa$ is a tunable parameter balancing exploitation and exploration when constructing the surrogate model. We take $\kappa=2.5$ as a default value in the model. Combining GPR and the acquisition function, the surrogate model can be constructed to approximate the minimum value in the design space. In our case, such BO methods are applied to obtain active surface typologies with minimal residual bacterial cells. For both material optimizations, we randomly explored the design space for 5000 material evaluations and repeated 5 times of different random seeds.

\clearpage
 
\begin{center}
    \textsf{\bfseries\Large Metaheuristic Optimization}
\end{center}
\section*{Methodology Generalization}

Although different methods may employ totally different notations and problem representations varied by their different optimization techniques, the general metaheuristic methods can be generalized to a ``heuristic sampling'' technique to sample the design space, in a non-rigorous sense. To symbolize our thoughts, assuming we have a design space map our design variables (recall we parametrize the elemental graph in Equation (\ref{eqn_opt_problem})) $\Theta$ to our objective values, one can write\begin{equation}
    DS: \Theta \rightarrow \mathcal{J}, \quad \Theta \in [\Theta_d^{low}, \Theta_d^{high}]
\end{equation}where $\Theta_d^{low}$ and $\Theta_d^{high}$ represent the lower and upper bounds of the design variables and $d$ indicate that the problem is $d$-dimensional. 

We generalize metaheuristic optimizations as sampling a set of points (or particles) with $d$ dimension, of which locations are represented as $X^d_1, X^d_2, ..., X^d_i, ..., X^d_m $ if we initialize the search with $m$ particles. This is how we perceive the metaheuristic optimization process shown in Equation (\ref{eqn_meta_update}), where we use $\mathcal{R}$ to denote the update strategy differed by different metaheuristic methods employing different prior information. Note that we here just propose a way to simplify the representation, but not state we will use this notation for all the further methods.

Here, we roughly classified our employed metaheuristic optimization methods into two categories, Nature-inspired algorithms, and mathematics-inspired algorithms. Of course, one may further categorize simulated annealing as physics-inspired (or materials-inspired), and genetic algorithm, particle swarm optimization, etc., as biology-inspired algorithms. For the sake of simplicity and clarity, we do not further proceed. In most implementations, we pick the default hyperparameters from the library or previous literature hoping to eliminate the bias towards hyperparameters.

\vspace{5pt}

\begin{center}
    \textsf{\bfseries\Large Nature Inspired Algorithms}
\end{center}

\section*{Genetic Algorithm}




Genetic algorithms have been recently applied to elemental composition design with rapid growth. The genetic algorithm was originally proposed inspired by the mutation and reproduction of genes. The algorithm seeks to mimic the fundamental mathematical rule to be treated as an optimization process. The process involves representing the design variables, denoted as $\theta$, as a set of "chromosomes," denoted as $\mathbb{C}$. The chromosomes can be written as $\mathbb{C} = [n_{atom}, \xi_n, \eta]$, or $\mathbb{C}\equiv[\Theta]$.

The population, denoted as $N_{pop}$, is selected as a randomly chosen assortment of chromosomes, such that $N_{pop} = [\mathbb{C}_1, \mathbb{C}_2, ...]$. For instance, if the population contains five chromosomes, it can be represented as $[10011]$. The population then undergoes "crossover" and "mutation" operations to generate the next generation of chromosomes, to maximize or minimize a given "fitness function," denoted as $\mathcal{J}$.

The probability of selecting a chromosome for reproduction is determined by its fitness value. Specifically, the probability of choosing the $ij^{\rm th}$ chromosome, denoted as $\mathbb{C}_{ij}$, is given by:\begin{equation}
P(\mathbb{C}_{ij}) = \left|\frac{f(\mathbb{C}_{ij})}{\sum_{k=1}^{N_{pop}}f(\mathbb{C}_k)}\right|,
\end{equation}where $f(\mathbb{C}_{ij})$ represents the fitness value of the chromosome, and $N_{pop}$ represents the total number of chromosomes in the population. Here, $f(\mathbb{C}_{ij})\equiv MEGNet(\Theta)$, where the MEGNet material property evaluations are defined as the fitness function(s). In the context of our generalization, one can treat each chromosome as a particle $X_i^D$, and $N_{pop}$ as the initial set of particles. The mutation and selection of chromosomes are analogous to the update of these particles and the hyperparameters in the update together with the probability determine the $\mathcal{R}$.

By iteratively applying crossover and mutation operations on the population and selecting chromosomes based on their fitness values, the genetic algorithm is able to identify the most promising set of design variables that maximize or minimize the fitness function. For the hyperparameters, for the single and multi-element cases, we pick the maximum iteration number as 25 and 75, respectively, and population size as 10, mutation probability as 0.3, crossover probability as 0.5, parents portion as 0.5, and uniform crossover type.

\section*{Particle Swarm Optimization}


Particle Swarm Optimization (PSO) is another evolutionary optimization algorithm that can be used for materials design. PSO starts with a set of randomized design variables (similar to chromosomes in GA) and propagates these points according to certain rules. The velocities and positions of these points in the $d^{th}$ dimension of the $i^{th}$ particle are denoted as $V$ and $X$ respectively. The update of these variables is given by:
\begin{equation}
        \begin{aligned}
            V_i^d &\leftarrow V_i^d + c_1 {\tt rand1}_i^d ({\tt pbest}_i^d - X_i^d) + c_2 {\tt rand2}_i^d ({\tt gbest}^d - X_i^d)\\
            X_i^d &\leftarrow X_i^d + V_i^d
        \end{aligned}
    \end{equation}

Here, ${\tt rand1}_i^d$ and ${\tt rand2}_i^d$ are random numbers between 0 and 1, $c_1$ and $c_2$ are acceleration coefficients, and ${\tt pbest}_i^d$ and ${\tt gbest}^d$ are the best positions of the $i^{th}$ particle and the swarm respectively in the $d^{th}$ dimension.

Liang et al.\cite{PSO_optimization} proposed a new formulation to improve traditional PSO by changing the update of velocities. The new formulation is given by:
\begin{equation}
        V_i^d \leftarrow wV_i^d + c_{local} {\tt rand}_i^d ({\tt pbest}_{fi(d)}^d - X_i^d) 
    \end{equation}

Here, $\mathbf{f}i = [f_i(1),f_i(2),f_i(3)]$ defines which particles' ${\tt pbest}$ the particle $i$ should follow. Additionally, $c_{local}$ is a local acceleration coefficient, $w$ is the inertia weight, and $flag_i$ represents the number of generations that the $i^{th}$ particle has not improved over its ${\tt pbest}$ position. In our PSO implementations, for the single and multi-element design cases, the population size is 10, $c_{local}$ equals 1.2, the parameter $flag_i$ equals 7, the range of the weight is $w\in[0.4,0.9]$, the portion of parents is selected as 0.5.



\section*{Hybrid Grey Wolf Optimization}

Grey wolf optimization (GWO) is a swarm intelligence algorithm that is inspired by the dynamics of hunting grey wolfs originally proposed by Mirjalili et al.\cite{gwo_original_optimization}. To explore the design space, GWO starts with a group of particles representing initial designs, denoted as $X_i=[n_{atom},\xi_n, \eta]$ for the $i^{th}$ design, with the position represented by $X$ (Equation (\ref{eqn_meta_update})). The GWO algorithm updates the population inspired by the hunting behavior of grey wolves. The algorithm comprises a pack of grey wolves, each representing a solution to the optimization problem. The original GWO algorithm updates the position of the $i^{th}$ wolf as follows:
\begin{equation}
X_i^d \leftarrow \frac{\left(X_{\alpha}^d - A_1 D_\alpha^d\right)+ \left(X_{\beta}^d - A_2 D_\beta^d\right) + \left(X_{\delta}^d - A_3 D_\delta^d\right)}{3}
\end{equation}where \begin{equation}
    \begin{aligned}
        D_\alpha = \left|C_1 X_\alpha^d - X_i^d\right|,\quad D_\beta = \left|C_2 X_\beta^d - X_i^d\right|,\quad D_\delta = \left|C_3 X_\delta^d - X_i^d\right|
    \end{aligned}
\end{equation}
where $X_i$ is the position of the $i^{th}$ wolf, $X_{\alpha}$, $x_{\beta}$ and $X_{\delta}$ are the positions of the $\alpha$, $\beta$, and $\delta$ wolves respectively, and $A_1$, $A_2$, $A_3$ are random coefficients. Here, both the coefficients ${\bf A} = [A_1,A_2,A_3]$ and ${\bf C}= [C_1,C_2,C_3]$ are calculated as\begin{equation}
    \begin{aligned}
        \mathbf{A} &= 2\mathbf{a}\mathbf{r}_1 - \mathbf{a}\\
        \mathbf{C} &= 2\mathbf{r}_2
    \end{aligned}
\end{equation}where components of $\bf a$ are linearly decreased from 2 to 0 during the material evaluations and $\mathbf{r}_1$ and $\mathbf{r}_2$ are random vectors in $[0,1]$. These randomization settings make the GWO a pseudo-hyper-parameter-free optimization method, as one does not need to set hyperparameters to update evaluations.

However, while the original GWO effectively explores potential areas in a large search space, its exploitation capability can be improved. To address this, Obadina et al.\cite{GWO_optimization} proposed combining GWO with the whale optimization algorithm (WOA) to use the logarithmic spiral equation from WOA to enhance GWO’s exploitation capability. This hybrid algorithm, denoted as GWO-WOA, updates the position of the $i^{th}$ wolf as the same as the original GWO, where the form of the two parameters $D_\alpha$ and $\bf a$ are different. For parameters $D_\alpha$ of the alpha wolf:
\begin{equation}
D_\alpha = \begin{cases}
    \mathcal{Y}\left|C_1 X_\alpha^d - X_i^d\right|,\quad {\rm if}\ p_1 <0.5\\\rho e^{bl}\cos(2\pi l)\left|C_1X_\alpha^d - X_i^d\right|,\quad{\rm if}\ p_1\geq 0.5
\end{cases}
\end{equation}
where $p_1$, $\mathcal{Y}$, and $\rho$ are random numbers within $[0,1]$. $l$ is a random number in $[-1,1]$. $b$ is a constant that defines the logarithmic spiral shape in WOA, which in our cases equals 1. As we mentioned in the vanilla GWO settings, $\bf a$ linearly decreased from 2 to 0, which can be written as ${\bf a} = 2-\frac{2{\tt iter}}{{\tt iter}_{max}}$, where {\tt iter} is the iteration number and ${\tt iter}_{max}$ is the maximum iterations. In the hybrid GWO (or GWO-WOA), this $\bf a$ is modified to\begin{equation}
    {\bf a} = 2\left(1 - \frac{{\tt iter}}{2{\tt iter}_{max}}\right)
\end{equation}where for now \textbf{a} decreases from 2 to 1. This hybrid GWO method, given a fixed $b=1$, can be considered a hyperparameter-free method. In our implementation, we set the population size as 10 and pick epochs as 10.

\section*{Hybrid-Improved Whale Optimization Algorithm}

Following our previous introduction to GWO, the Hybrid-Improved Whale Optimization Algorithm (HIWOA) is another optimization method that imitates the hunting behavior of blue whales searching for global optima. In the original WOA, the update depends on a random number $p$ and involves a linearly decreasing parameter $\bf a$ from 2 to 0 (similar to Hybrid GWO). The location of the whale (or particle) closest to the prey influences other whales' motion, in which the update writes:\begin{equation}
        \begin{aligned}
            {\rm if}\ p<0.5:\quad X_i^d &\leftarrow \ X^* - A \left| CX^* - X_i^d\right|\\{\rm where}:\quad A &= {\bf a}(2{\tt rand}_1 - 1)\\ C &= 2{\tt rand}_1
        \end{aligned}
    \end{equation}where $X^*$ represents the whale (or particle) that is closest to its prey. This update scheme is considered as the ``encircling prey''. ${\tt rand}_1$ is a random vector of $[0,1]$. When $p\geq 0.5$, the whale position update is considered as ``attacking prey'', where the update writes:\begin{equation}
        \begin{aligned}
            {\rm if}\ p\geq0.5:\quad X_i^d &\leftarrow X^* - D_1 e^{bl} \cos(2\pi l)\\
                {\rm where}:\quad D_1 &= \left|X^* - X_i^d\right|
        \end{aligned}
    \end{equation}where $b$ and $l$ holds the same definition as those of GWO. To improve global searchability, the whales surround prey when $|A|\leq 1$ or randomly select a whale position for reference when $|A|\geq 1$. This is usually referred as ``searching prey behavior'', for which the update of the whale position writes:\begin{equation}
        \begin{aligned}
            X_i^d \leftarrow X_{\tt rand} - A D_2\\
            {\rm where}:\quad D_2 = \left|CX_{\tt rand} - X_i^d\right|
        \end{aligned}
    \end{equation}where $X_{\tt rand}$ is a random whale location.


In the vanilla WOA, the update of the particle positions is strongly dependent on $\bf a$, where irrational randomized $\bf a$ values may lead to the failure of optimization. To overcome this, Tang et al.\cite{hiwoa_optimization} proposed the following update rules for HIWOA, in which they modify $\bf a$ as:\begin{equation}
    \mathbf{a} = 2 + 2\cos\left(\frac{\pi}{2}\times \left(1 + \frac{\tt iter}{{\tt iter}_{max}}\right)\right)
\end{equation}

Here, $w$ represents the inertia weight, modified by Tang et al.\cite{hiwoa_optimization} to the form: $w = 0.5 + 0.5 \left(\frac{{\tt iter}}{{\tt iter}_{max}}\right)^2$, which is a function of the iteration number. As the iteration number increases, the weight $w$ becomes larger, leading to a stronger dependence of the population on the best-so-far position $X^*$. Compared to the original WOA, HIWOA can improve the local search ability and achieve better optimization performance, in theory. In this implementation for materials discovery, the population size is set as 5, and the maximum feedback is also set to 5. 

\section*{Ant Colony Optimization}

    

Ant Colony Optimization (ACO) is similar to other population-based optimization algorithms such as PSO and WOA. It employs a population of virtual ants to search for the optimal solution, where each ant represents a candidate solution. The movement of each ant is determined by the difference between its position and the position of a randomly chosen ant, which is controlled by a step length parameter. Most importantly, the algorithm presents the dynamics (or motion) of the ant by mimicking the ant's foraging behavior by detecting the intensity of the pheromone in searching the food. Here, the food sources represent our design variables $\Theta$, where the ants explore the food sources to readjust the design variables to further tune the material properties. If assume there is a finite set of available solution components $\mathbf{C} = \{c_{ij}\}$, representing the candidate solution that the ant is currently evaluating. Associated with $c_{ij}$ is $\tau_{ij}$, is the pheromone trail intensity connecting the two ants $i$ and $j$. The probabilistic choice of solution components takes the form:\begin{equation}
    p(c_{ij}|s^p) = \frac{\tau_{ij}^\alpha \mathcal{W}(c_{ij})^\beta}{\sum_{c_{ij}\in N^p} \tau_{il}^\alpha \mathcal{W}(c_{il})^\beta}\label{candid}
\end{equation}where $\mathcal{W}(\cdot)$ is a weighting function assigns at each construction step a heuristic value to each solution candidate $c_{ij}$. $\alpha$ and $\beta$ are positive parameters determining the relation between pheromone and heuristic information. $N^p$ represents the feasible solution set, which can be interpreted as the analogy of the population in the genetic algorithm.


In addition to the movement of ants, ACO also updates a pheromone trail to facilitate the search process. The pheromone trail represents the quality of the solution at each position in the solution space. A higher concentration of pheromone indicates a better solution quality. The pheromone trail is updated based on the following equation:\begin{equation}
\tau_{ij}\leftarrow\begin{cases}
    (1 - \rho)\tau_{ij} + \rho\Delta\tau,\quad{\rm if}\ \tau_{ij} \in s_{good}\\
    (1 - \rho)\tau_{ij},\quad {\rm otherwise}
\end{cases}
\end{equation}where the pheromone evaporation rate is denoted by $\rho\in(0,1]$. $s_{good}$ is a chosen good solution, and $\Delta\tau$ is an associated interval to update $\tau$. ACO updates the pheromone to determine the optimal value between the defined ``food source'' and ``colony'', in which one can perceive the candidate solution $c_{ij}$ as the connection between the ant $X_i^d$ (in our formulation in Equation (\ref{eqn_meta_update})) and the design variables $\Theta$ in an analogous sense, and the pheromone $\tau_{ij}$ relate to the candidate solution $c_{ij}$ from the probabilistic choice (Equation (\ref{candid})). Here, the optimizations are initialized by counting 5 samples with a population size of 10.


\section*{Differential Evolution}


Differential evolution (DE) is very similar to GA, initializing the optimization set with a set of populations, and then conducting mutation, and selection, for iterative updates. The proposed algorithm commences with an initialization step, which involves generating the initial population of solutions. Recall our formulated problem in Equation (\ref{eqn_meta_update}), given a design optimization of $d$ dimensions, the $i^{th}$ particle's $j^{th}$ dimension ($j\in d$) can be initialized as:
\begin{equation}
X_{i,j}^0 = X_{j,L} + {\tt rand}(X_{j,U} - X_{j,L})
\end{equation}where $X_{j,L}$ and $X_{j,U}$ are the lower and upper bounds, respectively, for the $j$-th variable of the problem. $\tt rand$ returns to a random number in $[0,1]$.

After the initialization step, the mutation operation is performed on the population. The mutation is defined via a mutation vector that is constructed in various ways. One possible way is to use the following equation\cite{DE_optimization}:
\begin{equation}
\begin{aligned}
v_i^G =X_{r_1}^{G} + F (X_{r_2}^G-X_{r_3}^G),\quad r_1\neq r_2\neq r_3 \neq i
\end{aligned}
\end{equation}
where $r_1, r_2,$ and $r_3$ are randomly selected indices within the population (in the same concept as GA and ACO), distinct from each other and $i$, and $F$ is a scaling factor controls the amplification of the difference vector $(X_{r_2}^G-X_{r_3}^G)$.

The next step in the algorithm is the crossover operation, which is carried out via the binomial crossover. In this operation, the target vector is combined with the mutated vector to produce the trial vector, given by:
\begin{equation}
    u_{j,i}^G = \begin{cases}
        v_{j,i}^G,\ {\rm if}\ ({\tt rand}<Cr\ {\rm or}\ j = j_{rand})\\
        X_{j,i}^G,\ {\rm otherwise}
    \end{cases}
\end{equation}
where $j$ is the index of the current variable, ${\tt rand}$ is a random number between 0 and 1, $Cr$ is the crossover rate, and $j_{rand}$ is a randomly selected index.

Finally, the selection process is performed, in which the newly generated trial vector is compared with the parent vector of the same index, and the best of the two vectors is chosen as the parent for the next generation. This process can be expressed mathematically as:
\begin{equation}
    X_i^{G+1} = \begin{cases}
        u_i^G,\ MEGNet(u_i^G)\leq MEGNet(x_i^G)\\
        X_i^G, \ {\rm otherwise}
    \end{cases}
\end{equation}
where $MEGNet(\cdot)$ is the fitness evaluation function that measures the quality of each solution in the population, and $G$ is the generation number. In our DE implementation, we pick the weighting (scaling) factor $F$ as 0.7, the crossover rate is 0.9, and initialized with a population of 10. 

\section*{Simulated Annealing}

Simulated annealing (SA) is an optimization mimicking the annealing process of metals by defining a cooling process (i.e., decreasing of temperature) to perturb design variables by constructing a discrete-time inhomogeneous Markov chain to select the ``change'' of design variables and further determine the updates. The ``discrete-time'' characteristic can be represented by ``state'' (defined by epoch in the algorithm), symboled as $t$. Given our objective function $\mathcal{J}$ defined on a set $S$, and let $S^* \subset S$ be the best global optimum of $\mathcal{J}$. Additionally, we define a positive coefficient $q_{ij}$ ($j\in S(i)$) for every $i$, such that $\sum_{j\in S}q_{ij} = 1$\cite{sa_optimization}. 

To begin the simulated annealing process, we need an initial state $X(0) \in S$. Here, $X\equiv \Theta $ represents the vector of the design variables. The update of $X$ can be written as\cite{sa_optimization}\begin{equation}
        \begin{aligned}
           \text{if } \mathcal{J}(j)\leq \mathcal{J}(i):\quad &X(t) = j\\
 \text{if } \mathcal{J}(j)> \mathcal{J}(i):\quad &X(t) = j,\ {\rm with\ probability}\ {\tt exp}\left(-\frac{\mathcal{J}(j)-\mathcal{J}(i)}{T}\right)\\
 &X(t) = i,\ \rm otherwise
        \end{aligned}
    \end{equation}


Formally, the chosen of $X$ can be expressed as\begin{equation}
        P[X(t+1) = j | X(t)=i] = \begin{cases}
            q_{ij} {\tt exp}\left[- \frac{1}{T(t)} \max \{0, \mathcal{J}(j) - \mathcal{J}(i)\}\right],\ {\rm if}\ j\neq i, j \in S(i)\\
            0 ,\ {\rm if }\ j\neq i, j \not\in S(i)
        \end{cases}
    \end{equation}

This can be interpreted as a homogeneous Markov chain $X_T(t)$, where the temperature $T$ is held constant. The invariant probability distribution of the Markov chain $X_T$ is given by\begin{equation}
\pi_T(i) = \frac{1}{Z_T} {\tt exp}\left(- \frac{\mathcal{J}(i)}{T(t)}\right), \quad i \in S
\end{equation}where $Z_T$ is a normalizing constant. For the convergence of the algorithm, it can be shown that for any schedule $T(t) = \frac{d}{\log t}$  for all $t$: \begin{equation}
    \max_{X(0)} P\left[ X(t) \notin S^* | X(0) \right] \geq \frac{A}{t^\alpha}
\end{equation}where $A$ and $\alpha$ are positive constants depending on the function $\mathcal{J}$. A comprehensive detailed derivative can be found in Ref.\cite{sa_optimization}. In our implementations, we use the following hyperparameters: the population size is 10, the maximum number of sub-iteration is 1, the mutation rate is 0.1, the mutation step size is 0.1 and the damping factor for mutation step size is 0.99. For single and multi-element material optimization cases, the move counts per individual solution are 2 and 3, the epochs are 5 and 10, and the temperature changes are $500\rightarrow 1$ and $1000\rightarrow 1$, respectively.



\section*{Bees Optimization Algorithm}



The Bees Optimization Algorithm (BOA) is a bio-inspired optimization algorithm that shares similarities with other swarm intelligence algorithms such as PSO and GWO. It is inspired by the motion of bees. The group of artificial bees contains three sets: employed bees, onlookers, and scouts. the employed and onlooker bees are first played on ``food sources'' in the memory initially, and the scouts were sent to search for the new ``food sources'' for updates, where the food source can be treated as different sets of design variables $\Theta$. The update rule for BOA can be expressed as follows\cite{Bees2_optimization}:\begin{equation}
X_i^d \leftarrow X_i^d + \phi_{rand} (X_j - X_k)
\end{equation}where $X_i$, $X_j$, and $X_k$ are the current positions of three randomly selected bees from the population, and $\phi_{rand}$ is a random number between -1 and 1.

The bees in BOA select a food source based on the probability values of the food source, which are determined by:\begin{equation}
P_i = \frac{MEGNet_i}{\sum_{n=1}^N MEGNet_n}
\end{equation}

Here, $MEGNet_i$ is used as a fitness evaluation surrogate, which in the context of BOA represents the measure of the quality of the food source $i$ and $N$ is the total number of food sources in the population. In our BOA implementations, we use the following hyperparameters: the selected site ratio is 0.5, the elite site ratio is 0.2, the selected site bees ratio is 0.1, the elite site bees ratio is 1.2, the dancing radius is 0.1 and the dancing reduction is 0.99 with a population size of 10. We select 10 and 30 for single and elemental composition optimization cases, respectively.

\clearpage
 
\begin{center}
    \textsf{\bfseries\Large Mathematically Inspired Algorithms}
\end{center}


\section*{Arithmetic Optimization Algorithm}


The AOA algorithm is an optimization technique inspired by basic mathematical operations. The algorithm calculates two values, \texttt{MOA} and \texttt{MOP}, denote the math optimizer accelerated function and math optimizer probability, respectively, based on the current iteration and maximum iteration values. The \texttt{MOA} value is calculated as the minimum value plus the current iteration times the range of the search space divided by the maximum iteration. The \texttt{MOP} value is calculated as one minus the ratio of the current iteration raised to the power of 1 over a user-defined parameter $\alpha$, divided by the maximum iteration raised to the power of 1 over $\alpha$.

The update equations in the exploration and exploitation phases are as follows. MOA is a coefficient in the exploration phase to condition the use of different defined operators and MOP is a coefficient used in the update of the solution:\begin{equation}
    \begin{aligned}
        {\tt MOA}( {\tt iter})={\tt MOA}_{min} + {\tt iter}\left(\frac{{\tt MOA}_{max} - {\tt MOA}_{min}}{{\tt iter}_{max}}\right)\\
        {\tt MOP}({\tt iter}) = 1 - \frac{({\tt iter})^{1/\alpha}}{({\tt iter}_{max})^{1/\alpha}}
    \end{aligned}\end{equation}where $\tt iter$ is the current iteration and ${\tt iter}_{max}$ is the maximum iteration. ${\tt MOA}_{min}$ and ${\tt MOA}_{max}$ denote the minimum and maximum values of the accelerated function, respectively. $\alpha$ is a sensitivity parameter and is fixed to 5 in the implementation. The update of the solution, namely the exploration and exploitation stages, writes:
\begin{equation}
    \begin{aligned}
        X_{i,j}({\tt iter}+1)=\begin{cases}
            best(X_j) \div ({\tt MOP} +\epsilon) \times \left(\left(UB_j - LB_j\right)\times\mu + LB_j\right), & r_2<0.5\\
            best(X_j)\times {\tt MOP} \times \left(\left(UB_j - LB_j\right) \times \mu + LB_j\right), & otherwise
        \end{cases}\\
        X_{i,j}({\tt iter}+1)=\begin{cases}
            best(X_j) - {\tt MOP} \times\left(\left(UB_j - LB_j\right)\times\mu + LB_j\right), & r_3<0.5\\
            best(X_j) + {\tt MOP} \times\left(\left(UB_j - LB_j\right)\times \mu + LB_j\right), & otherwise
        \end{cases}
    \end{aligned}
\end{equation}where $X_{i,j}({\tt iter}+1)$ denotes the update of the $j^{th}$ position (in the population) of the $i^{th}$ solution. $best(X_j)$ is the best obtained solution of $j^{th}$ position. $UB_j$ and $LB_j$ are the upper and lower bound value of the $j^{th}$ position. $\mu$ is a control parameter in the search process and is fixed to 0.5. These equations update the positions of particles in the search space to improve the fitness value of the objective function being optimized. For the hyperparameters, we set $\alpha = 5$ and $\mu =0.5$. Here, ${\tt MOA}_{max}$ and ${\tt MOA}_{min}$ equals 0.9 and 0.2, respectively. The population size is 10. The epochs for single and multi-element composition optimizations are 10 and 30, respectively.


\section*{Runge Kutta Optimization}

The Runge Kutta Optimization (RUN) is an optimization method inspired by the Runge Kutta method for solving differential equations. The method employs the discretization strategy to search for the optimal points in a function distribution by employing the parameters $k_i$ to explore possible elemental compositions in our case. To begin with, recall first-order ODE with initial value problem (IVP):\begin{equation}
        \frac{dy}{dt} = f(x,y)\quad y(x_0)=y_0\label{rk_ode_eq}
    \end{equation}We can expand this equation on $d_t y$ (or can also be denoted as $\dot{y}$) based on Taylor series:\begin{equation}
        y(x+\Delta x) = y(x) + \dot{y}(x)\Delta x + \ddot{y}(x)\frac{(\Delta x)^2}{2!}+...\label{taylor_eq_ode}
    \end{equation}where $\Delta x$ is the interval. Equation (\ref{taylor_eq_ode}) can be discretized using various methods. The discretization for the differentiation based on the $4^{th}$ order Runge Kutta method (RK4) is written as:\begin{equation}
        f(x+\Delta x)= y(x) + \frac{1}{6}\left(k_1 + 2k_2 + 2k_3 + k_4\right)\Delta x
    \end{equation}in which the four weighted factors $(k_1, k_2, k_3, k_4)$ are given by: \begin{equation}
        \begin{aligned}
            k_1 = \dot{y}(x) = f(x,y)\\
            k_2 = f\left(x+ \frac{\Delta x}{2} , y + k_1 \frac{\Delta x}{2}\right)\\
            k_3 = f\left(x+ \frac{\Delta x}{2} , y + k_2 \frac{\Delta x}{2}\right)\\
            k_4 = f\left(x+\Delta x, y+k_3\Delta x\right)
        \end{aligned}
    \end{equation}where $k_1, k_2, k_3, k_4$ are the four increments that determine the slope of the intervals of different orders. This is the core ideology of Runge Kutta's discretization for numerically solving the ODE in Equation (\ref{rk_ode_eq}).
    
    Recall our formulated problem and analogy of using particles to sample the design space in Equation (\ref{eqn_meta_update}). Under the optimization, and inspired by the Runge Kutta discretization, the RUN optimization update of particle $i$ ($i$ in population $N^p$) with $d^{th}$ dimension follows:\begin{equation}
        X_i^d = L^d + {\tt rand}(U^d - L^d)
    \end{equation}where ${\tt rand}$ is a random number between $[0,1]$; $L^d$ and $U^d$ are the lower and upper bounds of the $d^{th}$ variable of the problem $(d=1,2,..., D)$ and here $D$ can be equal 2 or 3. $U^d$ and $L^d$ are thence set correspondingly given our discussion about Equation (\ref{eqn_opt_problem}).

    One of the key innovations and strengths, discussed in the main article, is that the RUN algorithm is hyperparameter-free, i.e., independent of the setting of hyperparameters. This is enabled by randomized parameters within the optimization process. The optimization algorithm can be initiated first by randomizing the four weighted factors parameters:\begin{equation}
        \begin{aligned}
            k_1 &= \frac{1}{2\Delta X}({\tt rand}\times X_w - u\times X_b),\quad u = {\tt round}(1+{\tt rand})\times (1-{\tt rand})\\
            k_2 &= \frac{1}{2\Delta X}({\tt rand} (X_w + {\tt rand}_1 k_1 \Delta x)-(uX_b + {\tt rand}_2 k_1 \Delta X))\\
            k_3 &= \frac{1}{2\Delta X}\left( {\tt rand} \left(X_w + {\tt rand}_1 \left(\frac{1}{2}k_2\right)\Delta X\right) - \left(uX_b + {\tt rand}_2 \left(\frac{1}{2}k_2\right)\Delta X\right)\right)\\
            k_4 &= \frac{1}{\Delta X} ({\tt rand} (X_w + {\tt rand}_1 k_3 \Delta X) - (uX_b + {\tt rand}_2k_3 \Delta X))
        \end{aligned}
    \end{equation}where $u$ is a random parameter to increase the importance of the best solution $X_b$. $X_w$ and $X_b$ are the worst and best solutions determined based on three random solutions $(X_{r_1}, X_{r_2}, X_{r_3})$ from the population, are determined from\begin{equation}
        \begin{aligned}
            \begin{cases}
                X_b = X_i, X_w = X_{bi}, \ MEGNet(X_n)<MEGNet(X_{bi})\\
                X_b = X_{bi}, X_w=X_i,\ MEGNet(X_n)\geq MEGNet(X_{bi})
            \end{cases}
        \end{aligned}
    \end{equation}where $X_{bi}$ is the best random solution in the population. The update schemes follow the rule\begin{equation}
        X_{i+1} \leftarrow \begin{cases}
            X_c + SF\times SM + \mu\times X_s,\ {\tt rand}<0.5\\
            X_m + SF\times SM + \mu\times X_s,\ {\tt rand}\geq0.5
        \end{cases}
    \end{equation}where $\mu = 0.5 + 0.1 \times \mathcal{N}({\tt rand})$, in which $\mathcal{N}({\tt rand})$ is a random number with a normal distribution. The formulation for $X_s$ and $X_s'$ are written as:\begin{equation}
    \begin{aligned}
        X_s &= \mathcal{N}({\tt rand}) (X_m - X_c),\quad 
        X_s' &= \mathcal{N}({\tt rand}) (X_{r_1} - X_{r_2})
    \end{aligned}
\end{equation}and in which $X_m$ and $X_c$ are calculated as\begin{equation}
    \begin{aligned}
        X_c = \phi \times X_n + (1 - \phi ) \times X_{r_1}\\
        X_m = \phi \times X_{best} + (1 - \phi) \times X_{lbest}
    \end{aligned}
\end{equation}Here, $SF$ and $SM$ are given by\begin{equation}
    \begin{aligned}
        SM = \frac{1}{6} (x_{RK})\Delta x,\ {\rm where}\ x_{RK} = k_1 + 2\times k_2 + 2\times k_3 + k_4\\
        SF = 2(0.5 - {\tt rand})\times\mathcal{F},\ {\rm where}\ \mathcal{F} = a\times e^{-b \times {\tt rand}\times \left(\frac{i}{max_i}\right)}
    \end{aligned}
\end{equation}where $a$ and $b$ are two constant numbers. Herein is only a simplified expression, the full derivation can be checked in Ref.\cite{RUN_optimization}. Here, we treated this problem as a hyperparameter-free scenario.

\end{document}